\documentclass{ieeeaccess}
\usepackage{cite}
\usepackage{amsmath,amssymb,amsfonts}
\usepackage{amsthm}
\usepackage{mathtools}
\usepackage{braket}

\usepackage{caption}

\theoremstyle{definition}
\newtheorem{thm}{Theorem}

\newtheorem{lem}[thm]{Lemma}
\newtheorem{prblm}{Problem}

\usepackage{color}

\usepackage{algorithm}
\usepackage{algpseudocode}
\algnewcommand{\StateIndent}[2]{\Statex \hskip#1em #2}

\DeclareMathOperator*{\argmin}{arg\,min}

\usepackage[pdftex]{graphicx}
\def\BibTeX{{\rm B\kern-.05em{\sc i\kern-.025em b}\kern-.08em
    T\kern-.1667em\lower.7ex\hbox{E}\kern-.125emX}}
\begin{document}
\history{Date of publication xxxx 00, 0000, date of current version xxxx 00, 0000.}
\doi{10.1109/TQE.2020.DOI}

\title{Post-processing variationally scheduled quantum algorithm for constrained combinatorial optimization problems}
\author{\uppercase{Tatsuhiko Shirai}\authorrefmark{1}, \IEEEmembership{Member, IEEE},
\uppercase{and Nozomu Togawa\authorrefmark{1}},
\IEEEmembership{Member, IEEE}}
\address[1]{Department of Computer Science and Communications Engineering, Waseda University}

\markboth
{Author \headeretal: Preparation of Papers for IEEE Transactions on Quantum Engineering}
{Author \headeretal: Preparation of Papers for IEEE Transactions on Quantum Engineering}

\corresp{Corresponding author: Tatsuhiko Shirai (email: tatsuhiko.shirai@aoni.waseda.jp).}

\begin{abstract}
We propose a post-processing variationally scheduled quantum algorithm (pVSQA) for solving constrained combinatorial optimization problems (COPs).
COPs are typically transformed into ground-state search problems of the Ising model on a quantum annealer or gate-based quantum device.
Variational methods are used to find an optimal schedule function that leads to high-quality solutions in a short amount of time.
Post-processing techniques convert the output solutions of the quantum devices to satisfy the constraints of the COPs.
pVSQA combines the variational methods and the post-processing technique.
We obtain a sufficient condition for constrained COPs to apply pVSQA based on a greedy post-processing algorithm.
We apply the proposed method to two constrained NP-hard COPs: the graph partitioning problem and the quadratic knapsack problem.
pVSQA on a simulator shows that a small number of variational parameters is sufficient to achieve a (near-)optimal performance within a predetermined operation time.
Then building upon the simulator results, we implement pVSQA on a quantum annealer and a gate-based quantum device. 
The experimental results demonstrate the effectiveness of our proposed method.
\end{abstract}


\titlepgskip=-15pt

\maketitle

\section{Introduction}
\label{sec:introduction}
\PARstart{Q}{uantum} algorithms for solving combinatorial optimization problems (COPs) have attracted attention in both academia and industry.
COPs are typically transformed into the ground-state search problems of the Ising models on quantum devices.
Adiabatic quantum annealing is a representative quantum algorithm to search for the ground state by following an adiabatic path~\cite{kadowaki1998quantum, farhi2000quantum}.
The computational time is approximately proportional to the square of the inverse of the spectral gap~\cite{morita2007convergence}.
Because the spectral gap generally decreases with problem size, adiabatic quantum annealing is difficult to execute for large-scale problems.

Recent advances in quantum technology have realized quantum devices such as quantum annealers~\cite{johnson2011quantum, maezawa2019toward} and gate-based quantum devices~\cite{kordzanganeh2023benchmarking}.
These devices provide diverse platforms for executing quantum computation, but they are susceptible to various forms of noise, including coherent and incoherent noise.
Noise limits their applicability to quantum algorithms with low computational costs.
For example, an experiment on a quantum annealer with $2,000$ spins revealed noise effects within a short timescale of approximately $100 \mathrm{n s}$~\cite{king2022coherent}.
Additionally, the computational accuracy of gate-based quantum devices decreases as the number of spins increases~\cite{kordzanganeh2023benchmarking}.

\begin{figure*}[th]
\centering
\includegraphics[width=0.85\linewidth]{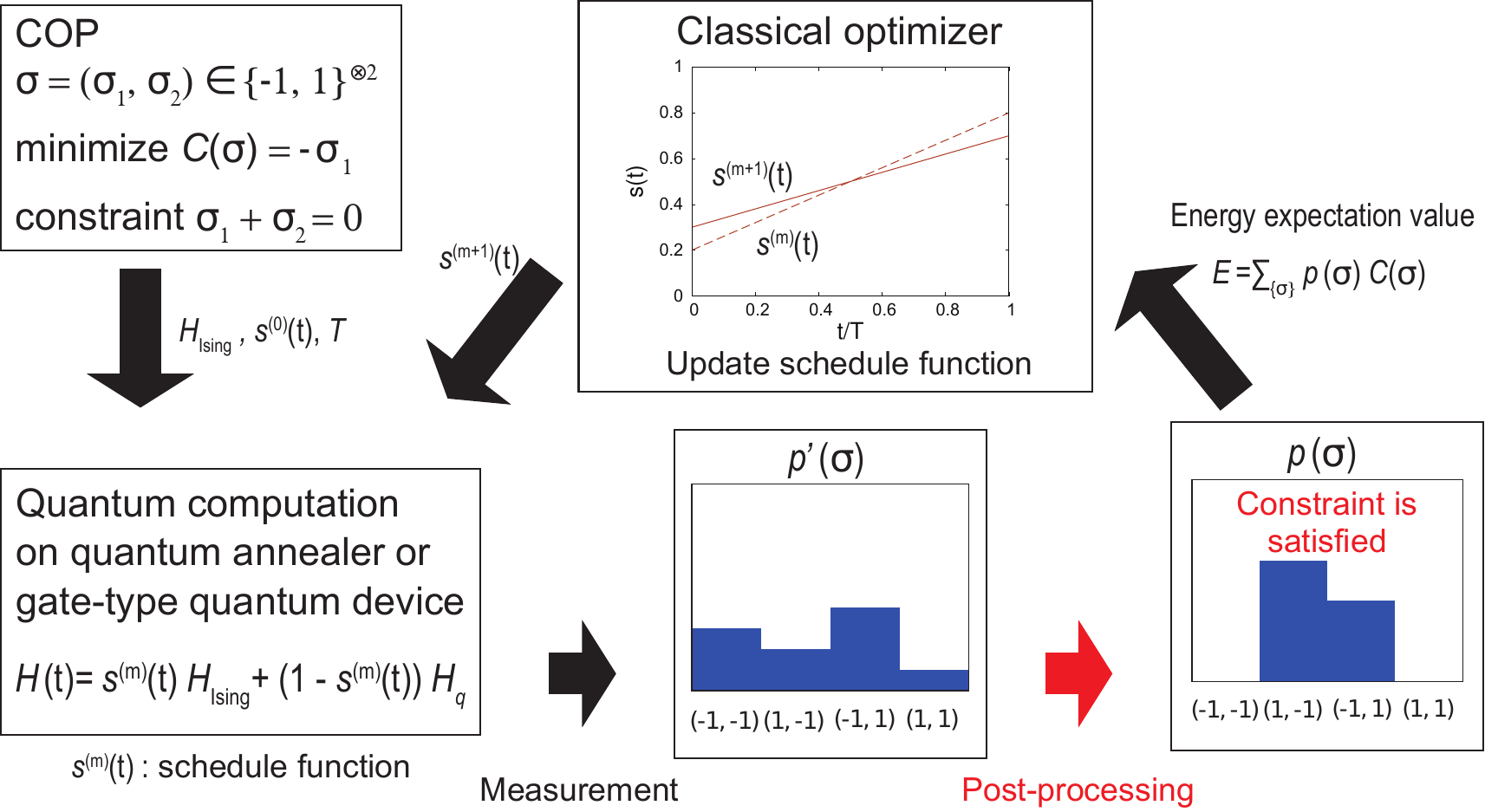}
\caption{
Illustration of pVSQA for a two-spin COP with a constraint $\sigma_1 + \sigma_2 = 0$.
Iterating the quantum computation and classical optimization $m$~times gives a schedule function $s^{(m)}(t)$.
$p'(\sigma)$ is the probability distribution of the spin configurations obtained by quantum measurements.
When a post-processing is performed, $p(\sigma)$ becomes populated only with feasible solutions.
The classical optimizer calculates the energy expectation value of the cost function $C(\sigma)$ using $p(\sigma)$ and updates the schedule function $s^{(m)}(t)$ (dashed line) to $s^{(m+1)}(t)$ (solid line). 
}
\label{Fig:schematic}
\end{figure*}

Given the severe limitations of the adiabatic quantum annealing and the near-term quantum devices, the development of quantum algorithms that obtain the low-energy states (i.e., high-quality solutions of COPs) within short operation time is crucial.
Several ideas have been proposed toward this direction.
One is to variationally optimize a schedule function to drive the system into the lower-energy state within a short amount of time (see Fig.~\ref{Fig:schematic} for an example of the schedule function $s^{(m)}(t)$ in this work).
Examples include variationally scheduled quantum simulation for quantum annealers~\cite{matsuura2021variationally} and quantum approximate optimization algorithm (QAOA) for gate-based quantum devices~\cite{farhi2014quantum}.
These methods can find a schedule function that leads to high-quality solutions within short operation time~\cite{zhou2020quantum, susa2021variational, cote2023diabatic, finvzgar2023designing}.

Another idea is to employ post-processing techniques.
When constrained COPs are considered, post-processing enhances the solution quality.
The spin configurations for a constrained COP are divided into two solutions: feasible and infeasible solutions~\cite{lucas2014Ising, shirai2022multi}.
Feasible solutions are those that satisfy the constraints.
Post-processing refers to a transformation of infeasible solutions into feasible ones.
Post-processing techniques have been applied to slot-placement problems and quadratic assignment problems~\cite{kuramata2021larger, fukada2021three}.

This study combines a variational method and a post-processing technique to efficiently obtain high-quality solutions of constrained COPs.
Fig.~\ref{Fig:schematic} provides a flowchart of our proposed algorithm, which we call the post-processing variationally scheduled quantum algorithm (pVSQA).
As an illustration, consider a COP with two spins $\sigma =(\sigma_1, \sigma_2) \in \{ -1, 1\}^{\otimes 2}$ subject to a constraint $\sigma_1 + \sigma_2 = 0$.
The spin configurations $(\sigma_1, \sigma_2)=(1,-1)$ and $(-1, 1)$ are feasible solutions, and all others are infeasible solutions.
First, the quantum device generates a variational quantum state via quantum computation.
Then quantum measurements of the quantum state provide the probability distribution of the spin configuration denoted by $p'(\sigma)$.
The probability distribution generally has a population over the infeasible solutions (i.e., $(\sigma_1, \sigma_2)=(-1,-1)$ and $(1, 1)$).
Post-processing transforms the infeasible solutions into feasible ones.
Hence, the probability distribution $p(\sigma)$ is populated only with feasible solutions.
The classical optimizer calculates the energy expectation value of the cost function using $p(\sigma)$ and updates the schedule function.
Repeating this process finds an optimized schedule function that gives low-energy states. 

The three main contributions of this paper are given as follows:
\begin{itemize}
    \item We propose pVSQA to efficiently provide high-quality solutions of constrained COPs.
    We obtain a sufficient condition for constrained COPs to apply pVSQA based on a greedy post-processing algorithm.
    We apply it to the graph partitioning problem (GPP) and the quadratic knapsack problem (QKP).
    \item We execute pVSQA on a simulator. pVSQA requires a small number of variational parameters to achieve a (near-)optimal performance within a predetermined operation time.
    \item The experiments on a quantum annealer and a gate-based quantum device demonstrates that pVSQA outperforms conventional quantum annealing with and without a post-processing and QAOA without a post-processing.
\end{itemize}

The rest of this paper is organized as follows.
Section~\ref{Sec:Related} provides the settings and preliminaries.
Section~\ref{Sec:Proposed} proposes pVSQA.
Section~\ref{Sec:Numerical} exhibits the simulator results. Section~\ref{Sec:Experiment} shows the experimental results on a quantum annealer and a gate-based quantum device.
Section~\ref{Sec:Discussion} discusses the results.
Section~\ref{Sec:Conclusion} summarizes this paper.

\section{Preliminaries}~\label{Sec:Related}
\hspace{-1mm}Ising models generically map various computationally hard COPs~\cite{lucas2014Ising,glover2018tutorial,shirai2023spin}.
Ising models are defined on undirected graph $G=(V,E)$ as
\begin{equation}
    \hat{H}_{\mathrm{Ising}} = \sum_{(i,j)\in E} J_{ij} \hat{\sigma}_i^z \hat{\sigma}_j^z + \sum_{i \in V} h_i \hat{\sigma}_i^z + H_0,
\end{equation}
where $\hat{\sigma}_i^{x,y,z}$ are the Pauli spins acting on site $i$, $J_{ij} \in \mathbb{R}$ denotes an interaction between spin~$i$ and spin~$j$, $h_i \in \mathbb{R}$ denotes a magnetic field on spin~$i$, and $H_0 \in \mathbb{R}$.
The Ising model for a constrained COP is given as
\begin{equation}
    \hat{H}_{\mathrm{Ising}} = \hat{H}_{\mathrm{obj}} + \hat{H}_{\mathrm{cst}},
\end{equation}
where $\hat{H}_{\mathrm{obj}}$ and $\hat{H}_{\mathrm{cst}}$ are Ising models for the objective function and the constraints of the COP, respectively.
The cost function $C(\sigma)$ is given as
\begin{equation}
    C(\sigma)=\bra{\sigma} \hat{H}_{\mathrm{obj}} \ket{\sigma}.
\end{equation}
Here, $\ket{\sigma}=\ket{\sigma_1} \otimes \ket{\sigma_2} \otimes \cdots \otimes \ket{\sigma_{|V|}}$ and $\ket{\sigma_i}$ for $i \in V$ is the eigenbasis of $\hat{\sigma}_i^z$.
Namely, $\hat{\sigma}_i^z \ket{\sigma_i} = \sigma_i \ket{\sigma_i}$ for $\sigma_i \in \{-1, 1\}$. 

Quadratic unconstrained binary optimization (QUBO) is equivalent to the Ising model~\cite{tanahashi2019application} and is defined on an undirected graph $G_q=(V_q, E_q)$ as
\begin{equation}
    Q = \sum_{(i,j) \in E_q} q_{ij} x_i x_j + \sum_{i \in V_q} q_{ii} x_i + Q_0,
\end{equation}
where $x_i \in \{ 0, 1\}$, $q_{ij} \in \mathbb{R}$, and $Q_0 \in \mathbb{R}$.
Replacing $x_i$ in the QUBO by $(\hat{\sigma}_i^z + 1)/2$ gives the Ising model.

This study considers the quantum computation of the following Hamiltonian dynamics  on a simulator or quantum devices, based on the quantum annealing algorithm and QAOA.
The quantum state $\ket{\psi (t)}$ obeys the Schr\"odinger equation:
\begin{equation}
    \frac{d}{dt} \ket{\psi (t)} = -i \left[s(t) \hat{H}_{\mathrm{Ising}} + (1-s(t)) \hat{H}_{\mathrm{q}}\right] \ket{\psi (t)},
    \label{shrodinger}
\end{equation}
where the Planck constant $\hbar=1$.
$\hat{H}_{\mathrm{q}}$ denotes the source of quantum fluctuations, and herein the transverse-field Hamiltonian $\hat{H}_{\mathrm{q}}=-\sum_{i\in V} \hat{\sigma}_i^x$ is adopted.
The initial state $\ket{\psi (0)}$ is set to the ground state of $\hat{H}_{\mathrm{q}}$.
The schedule function is characterized by $s(t)\in [0, 1]$ for $t\in [0, T]$, where the operation time is denoted by $T$.
$s(t)$ is variationally determined in pVSQA, whereas it is fixed in conventional quantum annealing.

\section{pVSQA}~\label{Sec:Proposed}
\hspace{-1.7mm}This section details pVSQA.
First, pVSQA is formulated.
Then we give three types of schedule functions since the function form depends on the type of quantum devices.
Finally, we give a condition for the applicability of pVSQA to constrained COPs.

\subsection{Formulation of pVSQA}
Fig.~\ref{Fig:schematic} illustrates the pVSQA.
The algorithm first gives the Ising model corresponding to a COP and the initial schedule function $s^{(0)}(t)$ for $t \in [0, T]$ as the input.
Then iterating the following process $m$ times updates the schedule function to $s^{(m)}(t)$.

The quantum device performs the quantum computation along the schedule function $s^{(m)}(t)$ to generate $\ket{\psi (T)}$.
The probability distribution of the spin configuration is given as
\begin{equation}
p'(\sigma) = \left| \langle \sigma \ket{\psi (T)} \right|^2.
\label{Eq:probability}
\end{equation}
When using a real quantum device, the probability distribution is approximately obtained by repeating the quantum projection measurements of $\ket{\psi(T)}$ on the computational basis.

Then a post-processing is performed.
The post-processing is defined by mapping $P$: $P\sigma \rightarrow \sigma$, which transforms all infeasible solutions into feasible ones (see Sec.~\ref{Sec:applicability} for the explicit mapping of $P$).
Furthermore, the probability distribution can be renormalized by using the top $r\%$ of solutions in terms of the cost function~\cite{barkoutsos2020improving}.
The probability distribution is recalculated as
\begin{equation}
p_r(\sigma) = \sum_{\sigma'|\sigma = P\sigma'} p'(\sigma') \frac{\theta(C_r-C(\sigma))}{r/100},
\label{Eq:modify_probability}
\end{equation}
where $\theta(\cdot)$ is the Heaviside step function and $C_r$ is chosen to satisfy the normalization of $p_r(\sigma)$ (i.e., $\sum_{\sigma} p_r(\sigma)=1$).
Here, $p_r(\sigma)$ is populated only with feasible solutions.
The energy expectation value of the cost function is given as
\begin{equation}
    E=\sum_{\sigma} p_r(\sigma) C(\sigma).
\end{equation}
A classical optimizer modifies the schedule function $s^{(m)}(t)$ to lower the energy expectation value and updates $s^{(m)}(t)$ to $s^{(m+1)}(t)$.
It relies on an optimization algorithm such as the gradient descent method or the Powell method to construct a progressively optimized sequence of variational parameters~\cite{dezvalle2021quantum}.

Finally, when the schedule function is optimized, the performance of pVSQA is evaluated using probability distribution $p_r(\sigma)$.

As a comparison to pVSQA, a variational quantum algorithm without a post-processing called VSQA is introduced.
The quantum device performs the quantum computation in \eqref{shrodinger}.
The probability distribution can be renormalized as
\begin{equation}
    p_r'(\sigma)=\sum_\sigma p'(\sigma) \frac{\theta(C'_r-C'(\sigma))}{r/100},
    \label{Eq:VSQA}
\end{equation}
where $0<r\leq 100$, $C'_r$ is a threshold to satisfy $\sum_\sigma p'_r(\sigma)=1$, and the cost function $C'(\sigma)$ is given as
\begin{equation}
    C'(\sigma)=\bra{\sigma} \hat{H}_{\mathrm{Ising}} \ket{\sigma}.
\end{equation}
A classical computer calculates the energy $E'$ as
\begin{equation}
    E'=\sum_{\sigma} p_r'(\sigma) C'(\sigma).
\end{equation}
Since the constraint Hamiltonian $H_{\mathrm{cst}}$ of the cost function increases the energy of infeasible solutions, feasible solutions are easily obtained.
Then a classical optimizer lowers $E'$ to update the schedule function $s^{(m)}(t)$.

Here, it is noted that pVSQA converges to the ground state of $\hat{H}_{\mathrm{Ising}}$ as $T$ tends to infinity.
The reasoning is as follows.
Let us recall that the adiabatic theorem ensures the convergence of quantum annealing~\cite{morita2007convergence}.
Since the schedule function of quantum annealing is a specific case of pVSQA, the optimal value of the cost function for pVSQA is no larger than that of quantum annealing.
Given that the cost function takes its minimum value at the ground state, it follows that pVSQA attains this state as well.

\subsection{Type of Schedule functions}\label{subsec:schedule}
We consider three types of schedule functions: continuous, linear, and QAOA schedules.
The continuous schedule represents $s(t)$ as a continuous function, the linear schedule represents $s(t)$ as a linear function, and the QAOA schedule has $s(t)$ that takes values of either zero or one.
Although the continuous schedule shows the best performance among the three, its implementation on quantum devices remains challenging.
On the other hand, the linear schedule can be performed on quantum annealers, whereas the QAOA schedule is easily implemented on gate-based quantum devices.

\begin{figure}[t]
\centering
\includegraphics[width=0.45\linewidth]{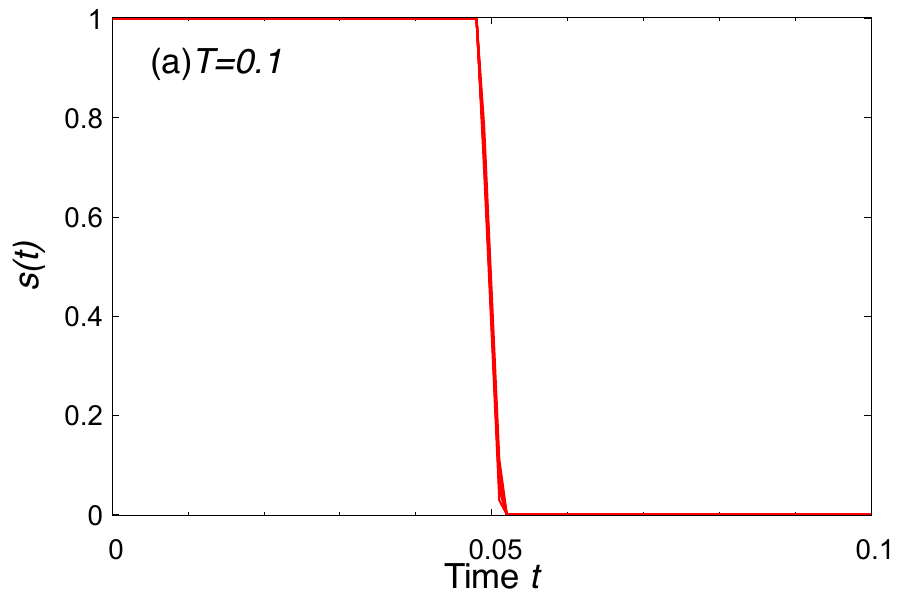}
\includegraphics[width=0.45\linewidth]{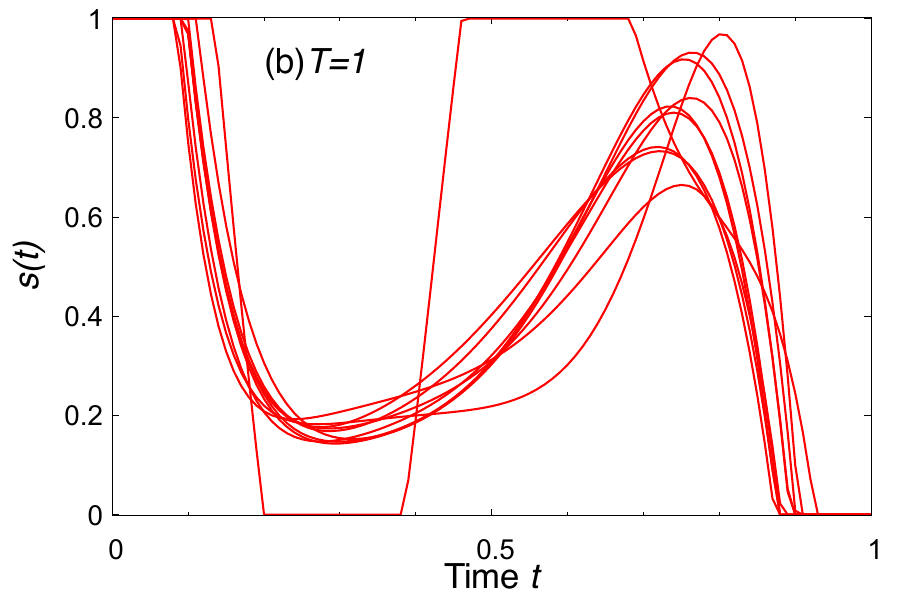}\\
\includegraphics[width=0.45\linewidth]{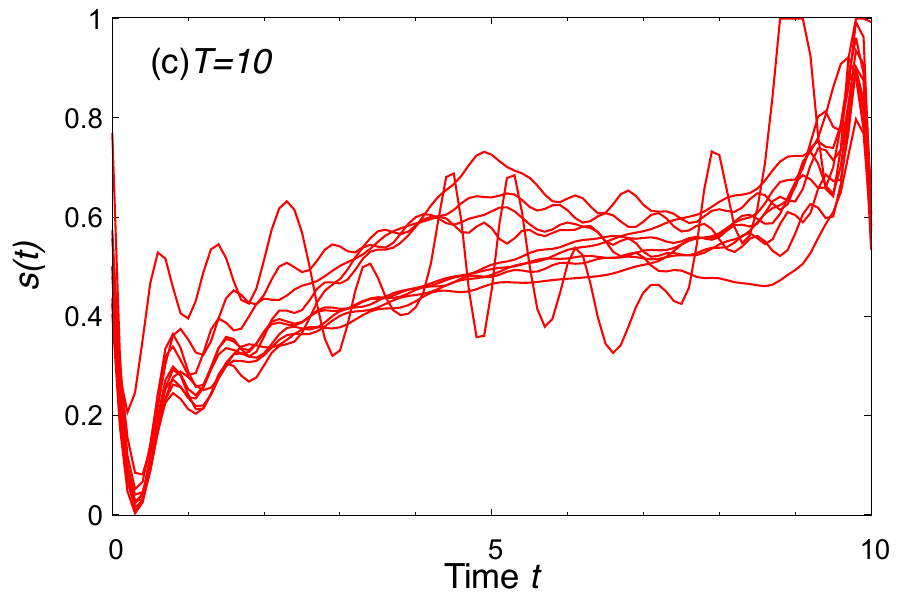}
\includegraphics[width=0.45\linewidth]{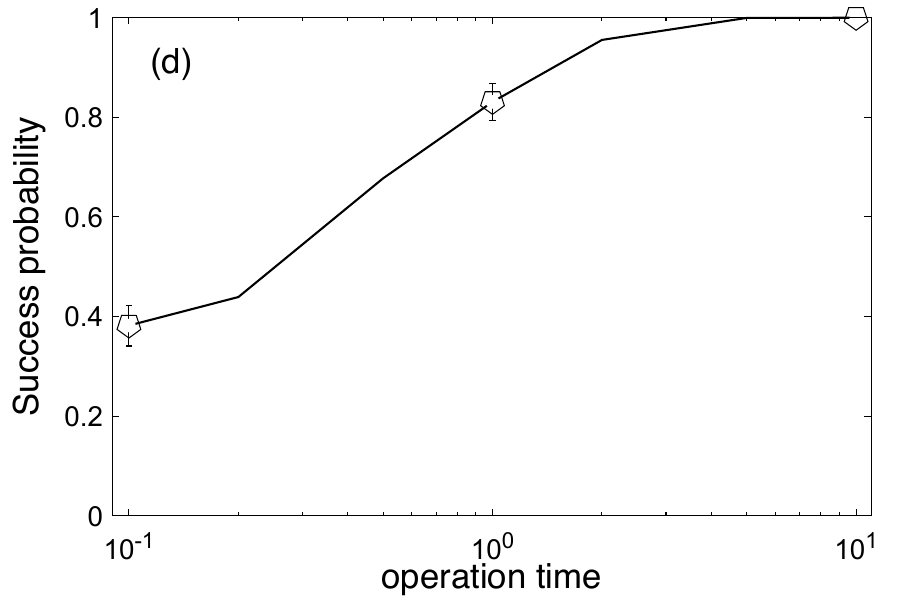}
\caption{
(a)--(c): Optimal schedule functions in the continuous schedule for different operation times of $T=0.1, 1$, and $10$, respectively.
Each line corresponds to the optimal schedule function for a different GPP.
(d): Success probability as a function of operation time.
Points in Fig.(a)--(c) are marked.}
\label{Fig:schedule}
\end{figure}

\begin{figure}[t]
\centering
\includegraphics[width=0.45\linewidth]{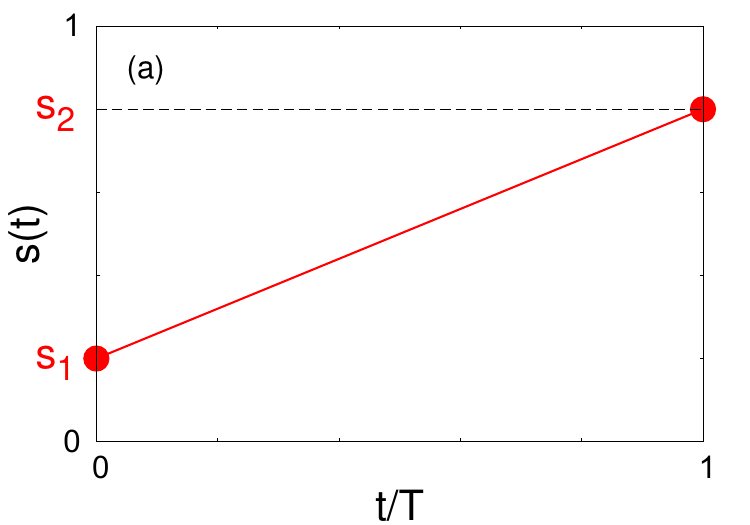}
\includegraphics[width=0.45\linewidth]{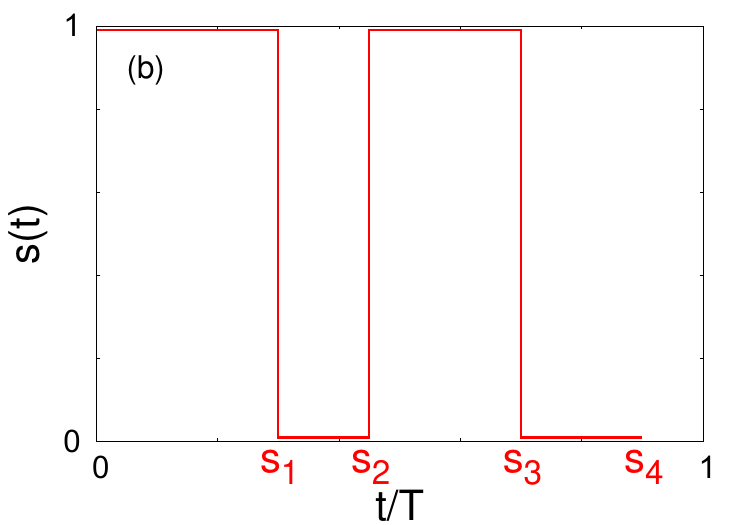}
\caption{
(a) Example of the linear schedule.
Variational parameters are given by $s_1$ and $s_2$.
(b) Example of the QAOA schedule with $2$ layers ($p=2$).
Variational parameters are given by $\{s_i\}_{i=1}^4$}
\label{Fig:linear_QAOA}
\end{figure}

First, we discuss the optimal schedule function $s^*(t)$ in the continuous schedule.
According to optimal control theory, for a short operation time $T$, $s^*(0)=1$ and $s^*(T)=0$, and the schedule function between the two can be a smooth path~\cite{brady2021optimal}.
The results can be extended straightforwardly to pVSQA with a post-processing.

Fig.~\ref{Fig:schedule}~(a)--(c) shows the three types of optimal schedule functions for $10$ GPPs with $8$ nodes for different values of the operation time.
Fig.~\ref{Fig:schedule}~(d) plots the success probability as a function of the operation time.
Here, the success probability denotes the probability of obtaining the optimum solution of a GPP when quantum computation is executed along the optimal schedule function and a subsequent post-processing without a renormalization (i.e., $r=100$ in \eqref{Eq:modify_probability}) is performed (see Sec.~\ref{Sec:applicability} for details).
The success probability increases with operation time.
Short operation times produce a bang-bang-type schedule function.
Quantum computation starts with the Ising Hamiltonian and quickly transitions to the transverse-field Hamiltonian.
The transition occurs at the midpoint of the operation time.
Intermediate operation times result in a bang-anneal-bang-type schedule function.
Quantum computation begins with the Ising Hamiltonian, which follows a continuous function that ends with the transverse-field Hamiltonian.
Long operation times generate schedule functions that closely resemble the conventional quantum annealing, which start with the transverse-field Hamiltonian and end with the Ising Hamiltonian.
Below, we call the short, intermediate, and long operation time regimes the bang-bang, bang-anneal-bang, and anneal regimes, respectively.
Reference~\cite{brady2021optimal} reported similar observation.

The linear schedule has two variational parameters $s_1$ and $s_2$, which represent the initial and final values of $s(t)$, respectively (see Fig.~\ref{Fig:linear_QAOA} (a)).
The schedule function is given as
\begin{equation}
    s(t)= s_1 + \frac{s_2-s_1}{T}t.
\end{equation}
It should be noted that previous studies fixed $s(0)=0$ and $s(T)=1$, and then used the values of $s(t)$ during quantum simulation as variational parameters~\cite{matsuura2021variationally, cote2023diabatic, finvzgar2023designing}. 
Unlike previous studies, we adopt $s_1$ and $s_2$ as variational parameters since the optimal schedule function in the continuous schedule does not always start with $s(0)=0$ or end with $s(T)=1$ (see Fig.~\ref{Fig:schedule}).
The linear schedule is approximately implemented on quantum annealers (see Sec.~\ref{subsubsec:annealer} for details).

The QAOA schedule has $p$ layers.
Each layer consists of two variational parameters (see Fig.~\ref{Fig:linear_QAOA} (b)).
Thus, there are $2p$ variational parameters denoted by $\{s_i\}_{i=1}^{2p}$.
The time duration of $s(t)=1$ and $s(t)=0$ for layer $\ell \in \{1, \ldots, p\}$ is respectively determined by
\begin{equation}
    s(t)= \left\{
    \begin{aligned}
        &1 \quad \text{ for }\quad s_{2\ell-2} \leq t < s_{2\ell-1}\\
        &0 \quad \text{ for }\quad s_{2\ell-1} \leq t < s_{2\ell},
    \end{aligned}
    \right.
    \label{Eq:qaoa}
\end{equation}
where $s_0=0$.
We impose the following conditions on the variational parameters: $s_{i-1} \leq s_i$ for $i\in \{1,\ldots, 2p\}$ and $s_{2p} \leq T$.
The quantum state is not time-evolved between $t \in [s_{2p}, T]$.
These conditions allow the performance of the QAOA schedule to be compared with the other types of schedules.
The QAOA schedule is easily implemented on real gate-based quantum devices (see Sec.~\ref{subsubsec:gate} for details).

\subsection{Applicability of pVSQA}\label{Sec:applicability}
We consider following COPs.
\begin{prblm}\label{problem1}
For binary variables $x_i \in \{0, 1\}$ for $i \in V$, find a configuration that minimizes the objective function $Q_{\mathrm{obj}}$ under the following constraints
\begin{equation}
b_{\mathrm{min}}^m \leq \sum_{i \in V^m} a_i^m x_i \leq b_{\mathrm{max}}^m \text{ for } m\in\{1,2, \ldots, M\},
\label{constraint1}
\end{equation}
\noindent where $V^m =\{ i \in V | a_i^m \neq 0\}$.
Here, $b_{\mathrm{min}}^m$, $b_{\mathrm{max}}^m$, and $a_i^m$ for $i\in V^m$ are integers and $M$ is the number of constraints.
\end{prblm}

Table~\ref{Table:class} classifies the constraints into two types: independent constraint and dependent constraint.
Constraints are called independent when $V^m \cap V^n = \varnothing$ for all pairs of $m$ and $n$ ($m \neq n$).
Otherwise, they are called dependent.

\begin{table}[t]
  \centering 
  \caption{Type of constraints.
   } 
  \scalebox{1}{
  \begin{tabular}{c|c}\hline
   Type & post-processing \\ \hline\hline
   Independent & Greedy\\ \hline
   Dependent & Adhoc \\
   \hline
   \end{tabular}
  }
  \label{Table:class}
\end{table}

pVSQA is applicable when post-processing transforms all infeasible solutions into feasible ones (i.e., mapping $P(\cdot)$ in \eqref{Eq:modify_probability} exists).
Here, we provide a sufficient condition to have a greedy post-processing algorithm.
The conditions hold for COPs with independent constraints and cover a wide range of COPs including GPPs, knapsack problems, and graph-coloring problems~\cite{lucas2014Ising}.
On the other hand, a post-processing algorithm should be considered for each COP with dependent constraints.
For example, an adhoc post-processing was found for quadratic assignment problems~\cite{kuramata2021larger}.
It is herein noted that a post-processing algorithm with a low computational cost is not found for all types of the constraints.
This situation occurs when finding a feasible solution is NP-complete.
Extending the applicability of pVSQA remains a challenge.

Below, we first prove a theorem for COPs with independent constraints.
Then we provide a greedy post-processing algorithm and demonstrate its applicability in two constrained COPs: GPP and QKP.

\subsubsection{Theorem}
First, in Problem~\ref{problem1}, we consider an objective function $Q_{\mathrm{obj}}$ with a single constraint (i.e., $M=1$):
\begin{equation}
b^1_{\mathrm{min}} \leq \sum_{i \in V^1} a^1_i x_i \leq b^1_{\mathrm{max}},
\label{constraint_single}
\end{equation}
where $b_{\mathrm{min}}^1$, $b_{\mathrm{max}}^1$, and $a_i^1$ for $i\in V^1 \subseteq V$ are integers.
Then we prove a following lemma.
\begin{lem}\label{lemma}
For Problem~\ref{problem1} with a constraint in \eqref{constraint_single}, consider a Hamiltonian $Q'$ as
\begin{equation}
    Q' = Q_{\mathrm{obj}} + A' Q'_{\mathrm{cst}},
    \label{eq:lemma}
\end{equation}
where $A'$ denotes a constraint coefficient and
\begin{equation}
    Q'_{\mathrm{cst}} = \max \left\{ 0, - b^1_{\mathrm{max}} + \sum_{i \in V^1} a_i^1 x_i, b^1_{\mathrm{min}} - \sum_{i \in V^1} a^1_i x_i \right\}.
\end{equation}
Then all local minimum solutions of $Q'$ are feasible solutions under the following conditions:
\begin{enumerate}
    \item $A' > \Delta Q_{\mathrm{obj}}^1$ where $\Delta Q_{\mathrm{obj}}^1$ is the maximum energy change of $Q_{\mathrm{obj}}$ when flipping a binary variable in $V^1$.
    \item $b_{\mathrm{max}}^1 - b_{\mathrm{min}}^1 \geq \max_{i \in V^1} |a^1_i| - 1$.
    \item Feasible solutions exist.
\end{enumerate}
Here, a flip indicates a change in the value of a binary variable from $0$ to $1$ or vice versa.
Then the local minimum solution is defined as a solution that does not lower the energy of $Q'$ by flipping any binary variable.
\end{lem}

\begin{proof}[Proof of Lemma~\ref{lemma}]
To prove the lemma, we show that all infeasible solutions are not local minimum solutions under the three conditions.
Suppose that $\{x_i \}_{i \in V}$ is an infeasible solution.
Then 
\begin{equation}
    b_{\min}^1 \geq \sum_{i\in V^1} a_i^1 x_i + 1
    \label{proof1}
\end{equation}
or
\begin{equation}
    b^1_{\max} \leq \sum_{i \in V^1} a^1_i x_i - 1
    \label{proof2}
\end{equation}
is satisfied.
When~\eqref{proof1} holds, we find from the second condition that for any $j \in V^1$
\begin{equation}
    \left( \sum_{i\in V^1} a^1_i x_i \right) + |a^1_j| \leq b_{\mathrm{min}} + |a^1_j| - 1 \leq b_{\mathrm{max}}.
    \label{proof3}
\end{equation}
This indicates that the configurations obtained by flipping a single binary variable do not satisfy~\eqref{proof2}.

Next, we show that the infeasible solution is not a local minimum solution using a proof by contradiction.
Assume that the infeasible solution is a local minimum solution.
Equation~\eqref{proof3} implies that if a flip increases the value of $\sum_{i\in V^1} a_i^1 x_i$, it lowers the value of $Q'$ when $A' > \Delta Q_{\mathrm{obj}}^1$.
Therefore, a flip should not increase the value of $\sum_{i\in V^1} a_i^1 x_i$, implying that $a^1_i \geq 0$ when $x_i = 1$ and $a^1_i \leq 0$ when $x_i = 0$.
Thus,
\begin{equation}
    b^1_{\mathrm{min}} > \sum_{i|a^1_i \geq 0} a^1_i \geq \sum_{i \in V^1} a^1_i x'_i
\end{equation}
for any configuration $\{x'_i\}$.
This contradicts the third condition that feasible solutions exist and indicates that $\{x_i\}$ is not a local minimum solution.
For infeasible solutions that satisfy \eqref{proof2}, similar arguments show that they are not local minimum solutions.
Consequently, when the three conditions are satisfied, all local minimum solutions are feasible solutions.
\end{proof}

Then Lemma~\ref{lemma} gives a following theorem.
\begin{thm}\label{theorem}
For Problem~\ref{problem1} with constraints in \eqref{constraint1}, consider a Hamiltonian $Q'$ as
\begin{equation}
    Q' = Q_{\mathrm{obj}} + A' \sum_{m=1}^M {Q'}^{m}_{\mathrm{cst}},
    \label{eq:theorem}
\end{equation}
where $A'$ denotes a constraint coefficient and
\begin{equation}
    {Q'}^m_{\mathrm{cst}} = \max \left\{ 0, - b^m_{\mathrm{max}} \mathalpha{+} \sum_{i \in V^m} a_i^m x_i, b^m_{\mathrm{min}} \mathalpha{-} \sum_{i \in V^m} a^m_i x_i \right\}.
\end{equation}
Then all local minimum solutions of $Q'$ are feasible solutions under the following conditions:
\begin{enumerate}
    \item $A' > \Delta Q_{\mathrm{obj}}$ where $\Delta Q_{\mathrm{obj}}$ is the maximum energy change of $Q_{\mathrm{obj}}$ when flipping a binary variable in $V$.
    \item $b_{\mathrm{max}}^m - b_{\mathrm{min}}^m \geq \max_{i \in V^m} |a^m_i| - 1$ for $m \in \{1, \ldots, M\}$.
    \item Constraints are independent (i.e., $V^m \cap V^n = \varnothing$ for all $m$ and $n$ $(m\neq n)$).
    \item Feasible solutions exist.
\end{enumerate}
\end{thm}

\begin{proof}[Proof of Theorem~\ref{theorem}]
We first show that all local minimum solutions of $Q'$ satisfy the $m$th constraint in \eqref{constraint1} under the four conditions of Theorem~\ref{theorem}.
Let us rewrite \eqref{eq:theorem} in the form of \eqref{eq:lemma} as
\begin{equation}
    Q' = \tilde{Q}_{\mathrm{obj}} + A' {Q'}^{m}_{\mathrm{cst}},
\end{equation}
where $\tilde{Q}_{\mathrm{obj}} = Q_{\mathrm{obj}} + A' \sum_{n=1 (n\neq m)}^M {Q'}^{n}_{\mathrm{cst}}$.
Then $\Delta \tilde{Q}_{\mathrm{obj}}^m$ and $\Delta Q_{\mathrm{obj}}^m$ are respectively introduced as the maximum energy change of $\tilde{Q}_{\mathrm{obj}}$ and $Q_{\mathrm{obj}}$ when flipping a binary variable in $V^m$.
The independent constraints indicate $\Delta \tilde{Q}_{\mathrm{obj}}^m = \Delta Q_{\mathrm{obj}}^m$, and then the first condition of Theorem~\ref{theorem} gives
\begin{equation}
    A' > \Delta Q_{\mathrm{obj}} \geq \Delta Q_{\mathrm{obj}}^m = \Delta \tilde{Q}_{\mathrm{obj}}^m.
\end{equation}
Thus, the first condition of Lemma~\ref{lemma} is satisfied.
The remaining conditions of Lemma~\ref{lemma} trivially hold from the conditions of Theorem~\ref{theorem}.
Lemma~\ref{lemma} implies that all local minimum solutions of $Q'$ satisfy the $m$th constraint in \eqref{constraint1}.

The above argument holds for any $m \in \{1, \ldots, M\}$.
Thus, all local minimum solutions of $Q'$ are feasible solutions under the four conditions of Theorem~\ref{theorem}.
\end{proof}

We give some remarks on Theorem~\ref{theorem}.
First, it includes the equality constraint called $k$-hot constraint.
An inequality constraint in \eqref{constraint1} with $b^m_{\mathrm{min}} = b^m_{\mathrm{max}} = k$, and $a^m_i = 1$ for $i \in V^m$ is reduced to the $k$-hot constraint:
\begin{equation}
\sum_{i \in V^m} x_i = k.
\label{constraint2}
\end{equation}
The second condition (i.e., $b_{\mathrm{max}}^m - b_{\mathrm{min}}^m \geq \max_{i \in V^m} |a^m_i| - 1$) always holds and the fourth condition implies $k \leq |V^m|$.
The constraint Hamiltonian for the $k$-hot constraint is given as
\begin{equation}
    {Q'}^m_{\mathrm{cst}} = \left|\sum_{i \in V^m} x_i -k \right|.
\end{equation}

Second, we comment on the conditions of Theorem~\ref{theorem}.
As for the first condition, when $Q_{\mathrm{obj}}$ is formulated as a QUBO (i.e., $Q_{\mathrm{obj}}=\sum_{(i,j)\in E_q}q_{ij} x_i x_j + \sum_{i \in V_q} q_i x_i$), an upper bound of $\Delta Q_{\mathrm{obj}}$ is given as
\begin{equation}
    \Delta Q_{\mathrm{obj}} \leq \max_{i \in V_q} \Big(|q_{ii}|+\sum_{j|(i,j)\in E_q} |q_{ij}|\Big).
\end{equation}
$A'$ can be set to a value larger than the right-hand side of the equation.
The second and third conditions are easily checked by looking into $\{ a_i^m \}_{i \in V^m}$, $b_{\mathrm{min}}^m$, and $b_{\mathrm{max}}^m$ in \eqref{constraint1}.
The fourth condition is checked by performing a local optimization of $Q'$  (e.g. Algorithm~\ref{Algo:greedy}) for a spin configuration.
Under the first, second, and third conditions, the spin configuration always reaches a feasible solution by repeating a local optimization when the fourth condition is satisfied.
Otherwise, the feasible solutions do not exist.

\subsubsection{Greedy post-processing algorithm}
\begin{algorithm}[t]
\caption{Greedy post-processing algorithm: $P(\sigma')=\sigma$ in \eqref{Eq:modify_probability}} \label{Algo:greedy}
\begin{algorithmic}[1]
\renewcommand{\algorithmicrequire}{\textbf{Input:}}
\Require $\sigma'$
\renewcommand{\algorithmicensure}{\textbf{Output:}}
\Ensure $\sigma$
\State $x_i \gets (\sigma'_i + 1)/2$ for $i \in V$
        \State $f \gets \mathsf{True}$
        \While{$f = \mathsf{True}$}
                \For{$i \in V$}
                        \State $\Delta Q_i \gets$ Energy change of $Q'$ when flipping $x_i$
                \EndFor
                \State $j \gets \min(\argmin_{i \in V} \Delta Q_i$)
                \If{$\Delta Q_j < 0$}
                        \State $x_j \gets 1-x_j$
                \Else
                        \State $f \gets \mathsf{False}$
                \EndIf
        \EndWhile
\State $\sigma_i \gets 2 x_i - 1$ for $i \in V$
\end{algorithmic}
\end{algorithm}

\begin{figure}[h]
\centering
\includegraphics[width=0.90\linewidth]{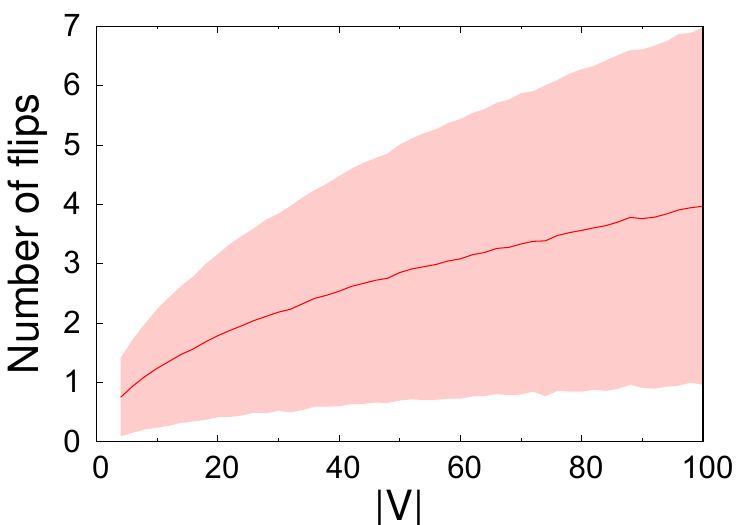}
\caption{
Number of flips to obtain a local minimum solution in the greedy post-processing algorithm as a function of $|V|$ for GPPs.
The solid line and shaded region represent the mean and the standard deviation over $1000$ random spin configurations for each of $10$ different GPPs, respectively.
}
\label{Fig:loop}
\end{figure}

Post-processing is given by a local optimization method using the property that all local minimum solutions are feasible solutions.
The local optimization method flips a binary variable to lower energy $Q'$.
Since infeasible solutions are not local minimum solutions, repeating the local optimization method can give a feasible solution.

In this study, we adopt a greedy method as a local optimization method.
Algorithm~\ref{Algo:greedy} gives a post-processing algorithm based on a greedy method.
The algorithm provides a mapping $P$ in pVSQA (see \eqref{Eq:modify_probability}).
The greedy method flips a binary variable resulting in the lowest energy of $Q'$ when it is flipped (line $4$--$12$).
We perform the greedy post-processing algorithm for all spin configurations obtained via the quantum computation.
Namely, if the obtained feasible solution is not a local minimum solution, post-processing transforms it into a feasible solution with a lower energy. 

Fig.~\ref{Fig:loop} shows the problem-size dependence of the number of flips to convert a spin configuration into a local minimum solution in the greedy post-processing algorithm for GPPs with a $|V|/2$-hot constraint.
The data represent the mean and the standard deviation over $1000$ random spin configurations for each of $10$ different GPPs.
The number of flips increases with the problem size, but the rate of increase slows down with the problem size.

\subsubsection{Examples of constrained COPs: GPP and QKP}
We consider two types of constrained COPs: GPP and QKP.
Both problems are categorized as NP-hard~\cite{stefan2012heuristic, pisinger2007quadratic}.

GPP divides the node set of a given graph $G_g = (V_g, E_g)$ into two subsets with equal size, minimizing the number of cuts.
Here, it is assumed that the graph has an even number of nodes.
The cut denotes the edge connecting the two subsets.
The binary variable $x_i \in \{0, 1\}$ is assigned for each node such that $x_i = 0$ ($x_i = 1$) indicates that node $i$ belongs to the one (the other) subset.
Then this problem has a $|V_g|/2$-hot constraint, which is given as
\begin{equation}
    \sum_{i \in V_g} x_i = \frac{|V_g|}{2}.
\end{equation}
pVSQA is applicable to GPP since the constraint satisfies the conditions of Theorem~\ref{theorem}.
QUBO for the GPP is given as~\cite{yoshimura2022qubo}
\begin{align}
    &Q_{\mathrm{gpp}} = Q_{\mathrm{obj}} + Q_{\mathrm{cst}}, \label{Eq:GPP}\\
    &\left\{
    \begin{aligned}
    Q_{\mathrm{obj}} &= -2 \sum_{(i,j)\in E_g} x_i x_j + \sum_{i \in V_g} k_i x_i,\nonumber\\
    Q_{\mathrm{cst}} &= A \left(\sum_{i \in V_g} x_i - \frac{|V_g|}{2}\right)^2,
    \end{aligned}
    \right.
\end{align}
where $k_i$ denotes the degree of node $i$. 
Here, $Q_{\mathrm{obj}}$ and $Q_{\mathrm{cst}}$ are QUBOs for the objective function and the constraints, respectively.
$A \geq 0$ is the constraint coefficient.

QKP finds a set of items to yield the highest profit within the capacity of the knapsack.
The set of $n$ items with weights $\{ w_i \}_{i=1}^n$ and profits $\{p_{ij}\}_{i \leq j}^n$, and the capacity of the knapsack $C$ are given.
Here, $p_{ii}$ denotes the profit of item~$i$ and $p_{ij}$ denotes the additional profit when item~$i$ and item~$j$ are chosen.
The binary variable $x_i \in \{0,1\}$ is assigned for each item such that $x_i = 1$ when item~$i$ is chosen and $x_i=0$ when item~$i$ is not chosen.
This problem has an inequality constraint, which is given as
\begin{equation}
    \sum_{i=1}^n w_i x_i \leq C.
\end{equation}
pVSQA is applicable to QKP since the constraint satisfies the conditions of Theorem~\ref{theorem}.
QUBO for the QKP is given as
\begin{align}
    &Q_{\mathrm{qkp}} = Q_{\mathrm{obj}} + Q_{\mathrm{cst}}, \label{Eq:QKP}\\
    &\left\{
    \begin{aligned}
    Q_{\mathrm{obj}} &= - \sum_{i=1}^n \sum_{j=i}^n p_{ij} x_i x_j,\nonumber\\
    Q_{\mathrm{cst}} &= A \left(\sum_{i = 1}^n \frac{w_i x_i}{C} \right)^2,
    \end{aligned}
    \right.
\end{align}
where $Q_{\mathrm{obj}}$ and $Q_{\mathrm{cst}}$ are QUBOs for the objective function and the constraints, respectively.
$A \geq 0$ is the constraint coefficient.
Here, we adopt a mapping in~\cite{mukasa2021scalable} for an inequality constraint.
Unlike other mappings~\cite{tamura2021performance}, it does not require additional auxiliary binary variables and is suitable for near-term quantum devices with severe limitations on the input number of spins.

\section{simulator results}~\label{Sec:Numerical}
\hspace{-1mm}The performance of pVSQA is initially evaluated using a simulator.
The simulator employs the fourth-order RungeKutta method to solve the Schr\"odinger equation in~\eqref{shrodinger}.
We use Python~3.7.6 integrated with intel Fortran as the implementation language.

\subsection{Instances of COPs}
We generate $10$ instances for GPPs and $10$ instances for QKPs for each problem size.
The GPP instances are specified by the undirected graph $G_g=(V_g,E_g)$.
Random graphs are created by connecting each pair of vertices with a probability of $0.5$ to form edges.
The QKP instances with $n$ items are generated by benchmarking instances with the $100$ items listed in~\cite{patvardhan2015solving}.
They are labeled $100\_100\_i$ where $i \in \{1, \ldots, 10\}$ denotes the instance.
Each instance is specified by a profit matrix $\{p^{(i)}_{jk}\}_{j\leq k}^{100}$, weight $\{ w^{(i)}_j \}_{j=1}^{100}$, and knapsack capacity $C^{(i)}_{100}$.
To generate QKPs with $n$ items, we use a profit matrix $\{p^{(i)}_{jk}\}_{j\leq k}^n$, weight $\{ w^{(i)}_j \}_{j=1}^n$, and knapsack capacity $C^{(i)}_{n}$, where $C^{(i)}_{n}=\lfloor nC^{(i)}_{100}/100 \rfloor$.
Here, $\lfloor x \rfloor \coloneqq \max \{ m \in \mathbb{Z}| m \leq x \}$ is the usual floor function.
The optimum solution for each problem is obtained by a state-of-the-art heuristic Ising-model classical solver~\cite{fixstars}.

For each problem instance, we calculate the Ising model using the QUBOs in~\eqref{Eq:GPP} and~\eqref{Eq:QKP}.
The coefficients of the Ising model are rescaled so that the maximal absolute value of $\{ J_{ij} \}_{(i,j)\in E}$ and $\{ h_i \}_{i \in V}$ is one.
This rescaling determines the timescale for quantum computation.

\subsection{Method}
We compare the performances of pVSQA, VSQA, pQA, and QA.
Here, pQA represents quantum annealing with a post-processing, and QA represents quantum annealing without a post-processing.
In pQA and QA, we adopt a linearly interpolated schedule function from $s(0)=0$ to $s(T)=1$ (i.e., $s(t)=t/T$).
pVSQA and VSQA optimize the schedule function using a classical optimizer. 
For continuous schedules, we follow the method in~\cite{brady2021optimal} based on optimal control theory and use a gradient descent method as a classical optimizer with an initial condition of $s(t)=0.5$ for $t \in [0, T]$.
We set $r=100$ in \eqref{Eq:modify_probability} and \eqref{Eq:VSQA}. 
For linear and QAOA schedules, we adopt the Powell optimizer implemented in SciPy as a classical optimizer~\cite{scipy}.
For the linear schedule, the maximum number of iterations is set to $10$ with initial values of $s_1=s_2=0.5$.
For the QAOA schedule, the maximum number of iterations is set to $10p$ and initially $s_i=0$ for $i \in \{1, \ldots, 2p\}$~\cite{wiersema2020exploring}.
The default values are used for the remaining input parameters.

The constraint coefficient $A$ is determined using the procedure in~\cite{tamura2021performance}.
Hereafter, $p_{100}(\sigma)$ and ${p}_{100}'(\sigma)$ respectively denote the probability distributions in~\eqref{Eq:modify_probability} and~\eqref{Eq:VSQA} obtained for either the optimized schedule function in pVSQA and VSQA or the given schedule function in pQA and QA.
We calculate the feasible solution rate and the average cost function.
The feasible solution rate is $1$ for pVSQA and pQA due to the post-processing.
The feasible solution rate for VQA and QA is calculated as
\begin{equation}
    p_{\mathrm{FS}} = \sum_{\sigma \in F} {p}_{100}'(\sigma),
\end{equation}
where $F$ is the feasible solution subspace.
The average cost function is given as
\begin{equation}
    C_{\mathrm{ave}} =\left\{
    \begin{aligned}
    &\sum_{\sigma} C(\sigma) p_{100}(\sigma) \text{ for pVSQA and pQA},\\
    &\frac{\sum_{\sigma \in F} C(\sigma) {p}_{100}'(\sigma)}{p_{\mathrm{FS}}} \text{ for VSQA and QA}.
    \end{aligned}
    \right.
\end{equation}
$C_{\mathrm{ave}}$ is calculated for the feasible solutions for VSQA and QA.
A smaller value of the average cost function indicates a higher performance.
To systematically determine the optimum value of the constraint coefficient, we set the threshold value of the feasible solution rate to $0.1$.
We repeatedly solve each problem instance while varying the constraint coefficient.
Then we determine the parameter regime that yields a feasible solution rate above the threshold value.
Finally, we search for the optimal value of $A$ to minimize the average cost function.
The precision threshold of the constraint coefficient is set to $1$ for GPPs and $200$ for QKPs.

The performances of pVSQA, VSQA, pQA, and QA are compared using the success probability averaged over the $10$ problem instances for each problem size.
The success probability $p_{\mathrm{suc}}$ is given as
\begin{equation}
    p_{\mathrm{suc}} =\left\{
    \begin{aligned}
    &\sum_{\sigma \in S} p_{100}(\sigma) \text{ for pVSQA and pQA},\\
    &\sum_{\sigma \in S} {p}_{100}'(\sigma) \text{ for VSQA and QA},
    \end{aligned}
    \right.
\end{equation}
where $S$ is the set of optimum solutions.

\subsection{Results}
\begin{figure}[t]
\centering
\includegraphics[width=0.8\linewidth]{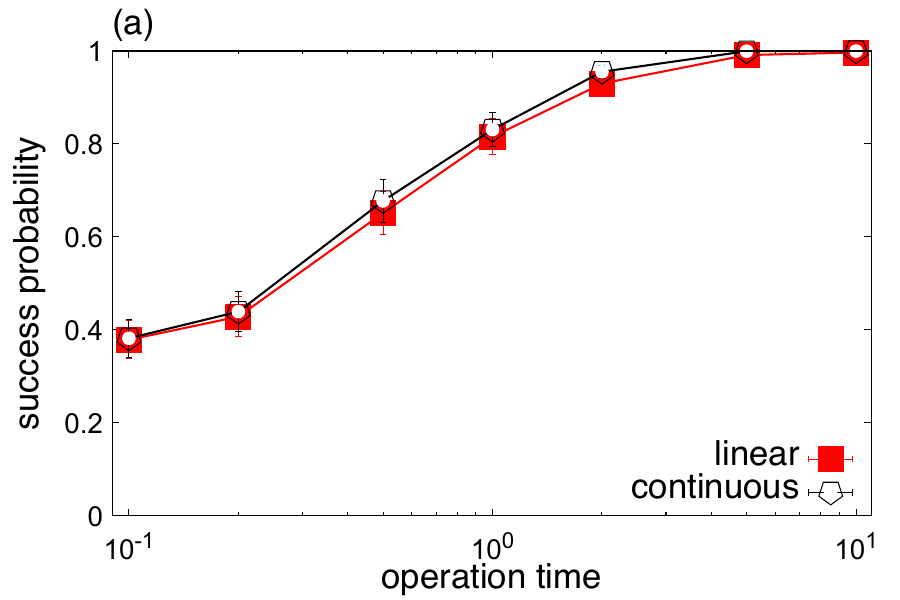}
\includegraphics[width=0.8\linewidth]{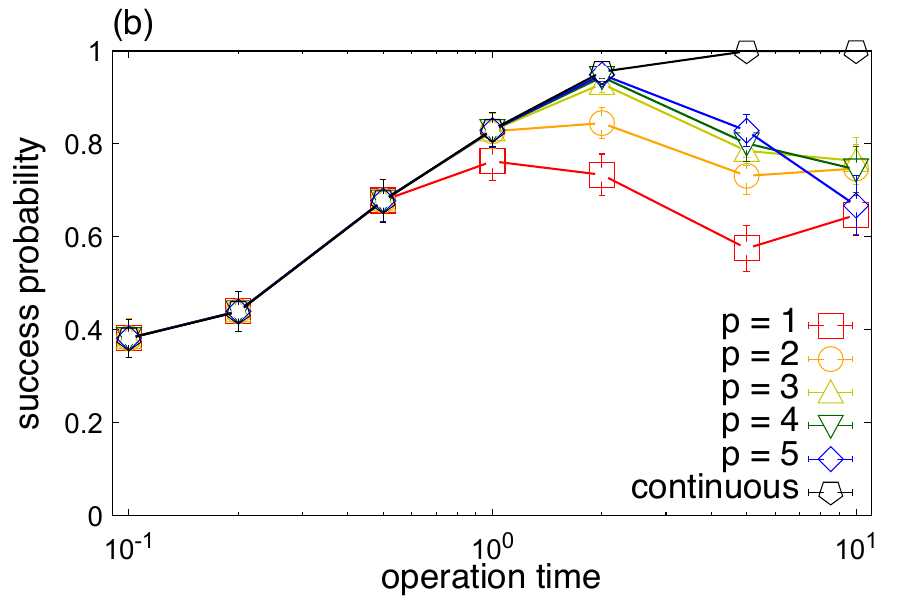}
\includegraphics[width=0.8\linewidth]{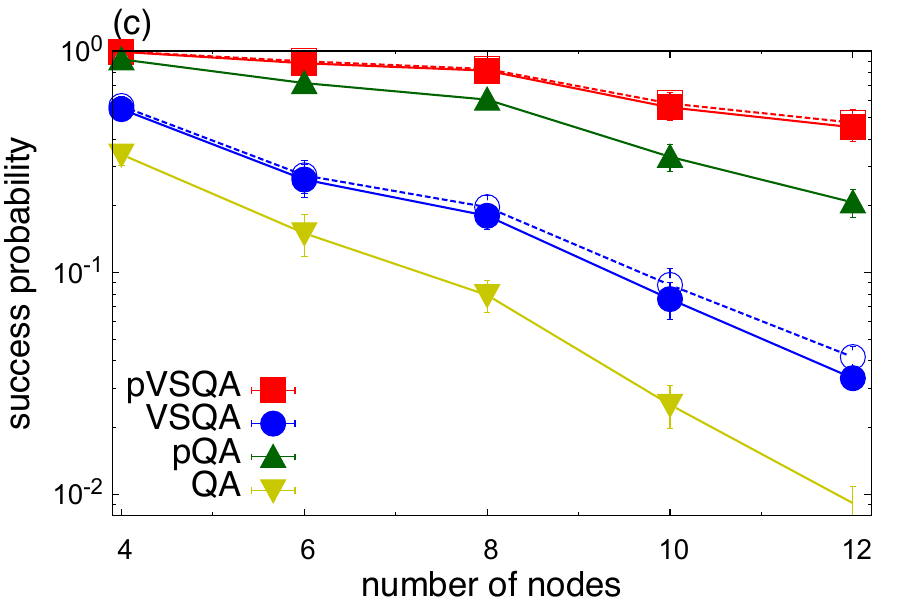}
\caption{
Results for GPPs.
Operation time dependence of the success probability of $8$ node GPPs for (a) the linear schedule and (b) the QAOA schedule.
Continuous schedule is depicted by pentagons (black) for comparison.
(c): Dependences of the success probability on the number of nodes for pVSQA, VSQA, pQA, and QA.
In pVSQA and VSQA, data for the linear schedule and the QAOA schedule are represented by the filled and open symbols, respectively.
}
\label{Fig:gpp}
\end{figure}

\begin{figure}[t]
\centering
\includegraphics[width=0.8\linewidth]{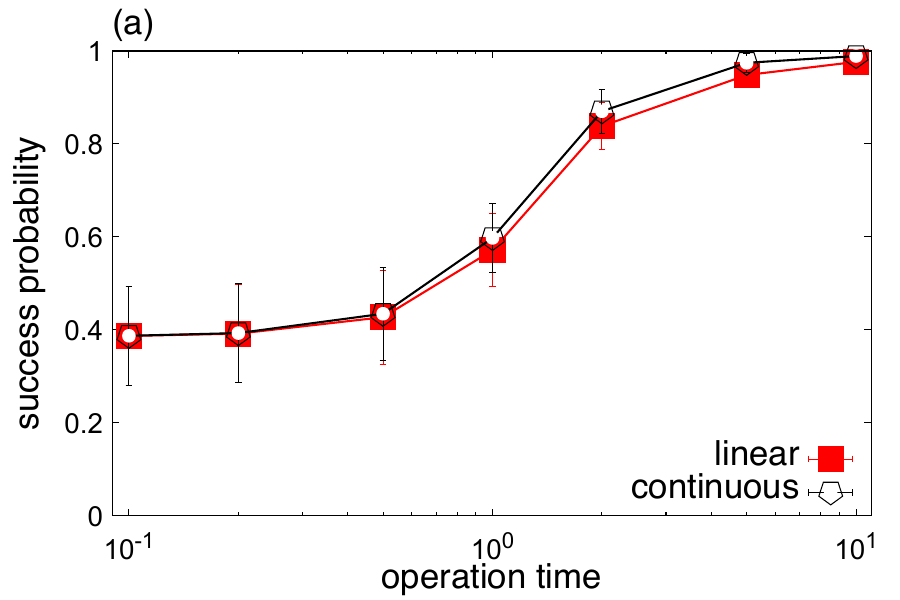}
\includegraphics[width=0.8\linewidth]{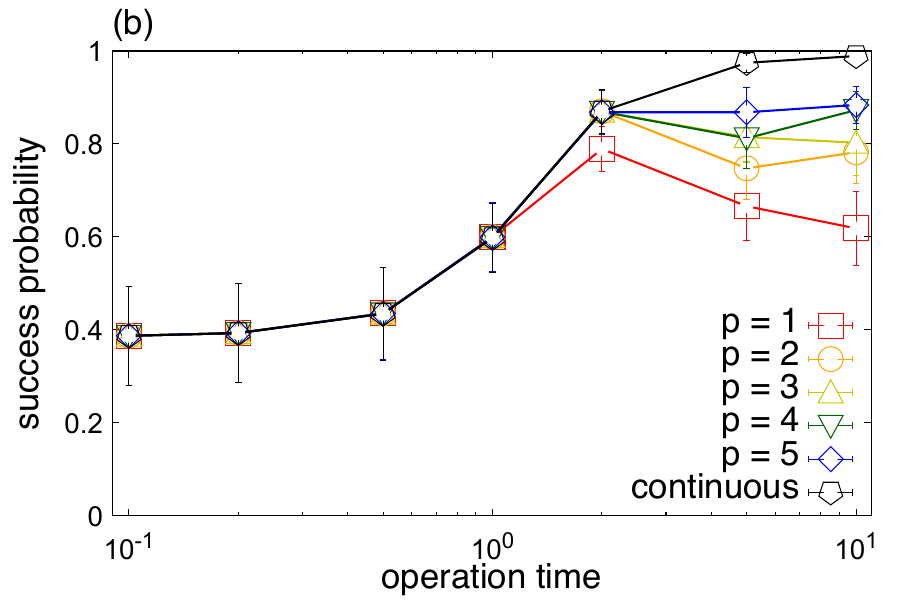}
\includegraphics[width=0.8\linewidth]{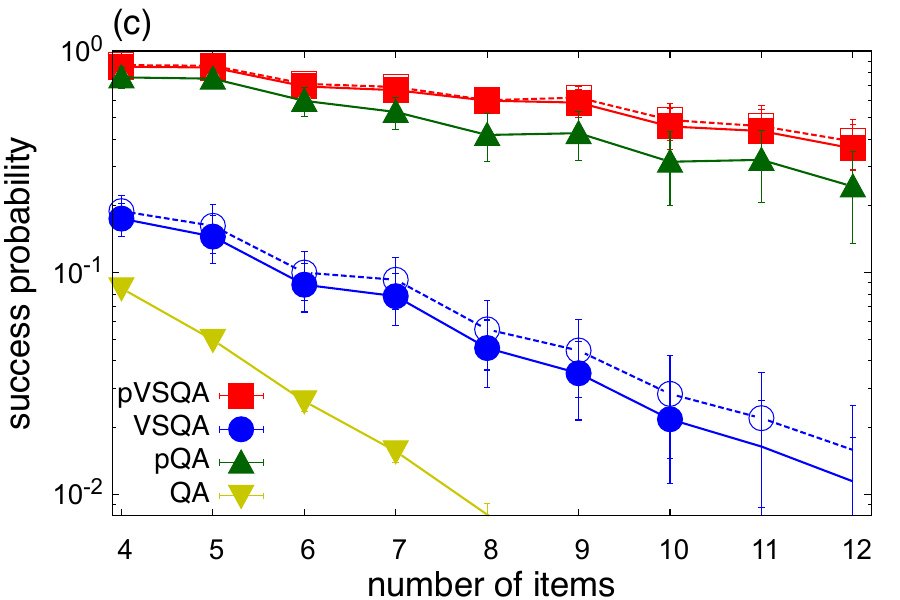}
\caption{
Results for QKPs.
Success probability of $8$ item QKPs as a function of operation time for (a) the linear schedule and (b) QAOA schedule.
Continuous schedule is depicted by pentagons (black) for comparison.
(c): Dependences of the success probability on the number of items for pVSQA, VSQA, pQA, and QA.
In pVSQA and VSQA, data for the linear schedule and QAOA schedule are represented by filled and open symbols, respectively.
}
\label{Fig:qkp}
\end{figure}

\hspace{-0mm}Fig.~\ref{Fig:gpp} shows the results for the GPPs.
The success probability for the linear schedule increases with the operation time.
The linear schedule almost reproduces the success probability of the continuous schedule for all operation times (Fig.~\ref{Fig:gpp} (a)).
For the QAOA schedule, $p=1$ is sufficient for the bang-bang regime ($T \sim 10^{-1}$), whereas $p=3$ is appropriate for the bang-anneal-bang regime ($T \sim 1$) to achieve the success probability of the continuous schedule (Fig.~\ref{Fig:gpp} (b)).
However, the performance of the QAOA schedule degrades in the anneal regime ($T \sim 10$) because the time steps in the QAOA schedule are too large to accurately approximate the continuous schedule.

Fig.~\ref{Fig:gpp} (c) compares the success probability of pVSQA with VSQA, pQA, and QA when the operation time is $T=1$ and the number of layers for the QAOA schedule is $p=3$.
Filled symbols connected by solid lines denote the results for the linear schedule, whereas open symbols connected by dashed lines indicate the results for the QAOA schedule.
For all the problem sizes, the linear and QAOA schedules exhibit almost the same performances.
pVSQA shows the best performance among them and significantly outperforms the conventional methods of QA and QAOA (i.e., VSQA for the QAOA schedule).

The simulator results for the QKPs are qualitatively similar to those for the GPPs (Fig.~\ref{Fig:qkp}).
The linear schedule approximates the continuous schedule well for all operation times (Fig.~\ref{Fig:qkp} (a)).
The QAOA schedule with a small value of $p$ provides a good approximation of the continuous schedule in the bang-bang regime and the bang-anneal-bang regime, but its performance degrades in the anneal regime (Fig.~\ref{Fig:qkp} (b)).
Fig.~\ref{Fig:qkp} (c) compares the success probability of pVSQA with the other methods.
Note that we set $p=3$ for the QAOA schedule and $T=1$.
pVSQA outperforms the other methods.

Table~\ref{Table:s1s2} shows the optimized values of variational parameters $s_1$ and $s_2$ for the linear schedule.
GPPs and QKPs display the similar dependences of $s_1$ and $s_2$ on the operation time.
For a short operation time, $s_1 \simeq 1$ and $s_2 \simeq 0$.
In contrast, $s_1 \simeq 0$ and $s_2 \simeq 1$ for a long operation time.
As the operation time increases, $s_1$ gradually decreases, whereas $s_2$ gradually increases.
These operation time dependences of $s_1$ and $s_2$ are consistent with the optimal schedule functions in the continuous schedule (see Fig.~\ref{Fig:schedule}).

\begin{table}[t]
  \centering 
  \caption{Optimized values of variational parameters $s_1$ and $s_2$ for the linear schedule.
  Values in parentheses denote the standard deviation of the average.
   } 
  \scalebox{1}{
  \begin{tabular}{c|cc|cc}\hline
    \hline
   \multicolumn{1}{c|}{} & \multicolumn{2}{c|}{GPP} & \multicolumn{2}{c}{QKP} \\ \hline
   $T$ & $s_1$ & $s_2$ & $s_1$ & $s_2$\\ \hline
   $0.1$ & $1.0 (0)$ & $0.0 (0)$ & $0.93 (5)$ & $0.14 (9)$ \\
   $0.2$ & $1.0 (0)$ & $0.0 (0)$ & $0.96 (3)$ & $0.16 (10)$ \\
   $0.5$ & $0.98 (1)$ & $0.0 (0)$ & $0.99 (1)$ & $0.17 (11)$ \\
   $1$ & $0.50 (2)$ & $0.35 (3)$ & $0.99 (1)$ & $0.16 (11)$ \\
   $2$ & $0.18 (3)$ & $0.67 (4)$ & $0.65 (5)$ & $0.48 (6)$ \\
   $5$ & $0.10 (2)$ & $0.79 (3)$ & $0.65 (5)$ & $0.48 (6)$ \\
   $10$ & $0.05 (2)$ & $0.84 (4)$ & $0.25 (12)$ & $0.92 (5)$ \\ \hline
   \end{tabular}
  }
  \label{Table:s1s2}
\end{table}

\section{Experiments with quantum devices}~\label{Sec:Experiment}
\hspace{-1mm}We conducted experiments of pVSQA on a quantum annealer and a gate-based quantum device.
We use a D-Wave annealer as a quantum annealer.
The quantum annealing hardware embeds a maximum of $180$ spins on a complete graph~\cite{boothby2020next}.
As a gate-based quantum device, we use {\it ibmq\_brisbane} and {\it ibmq\_osaka}, which are IBM Quantum Canary Processors~\cite{ibmq2021}.
The number of spins available is $127$.
Both experiments used Python~3.7.6 as the implementation language.

\subsection{Instances of COPs}
\begin{figure}[th]
\centering
\includegraphics[width=0.3\linewidth]{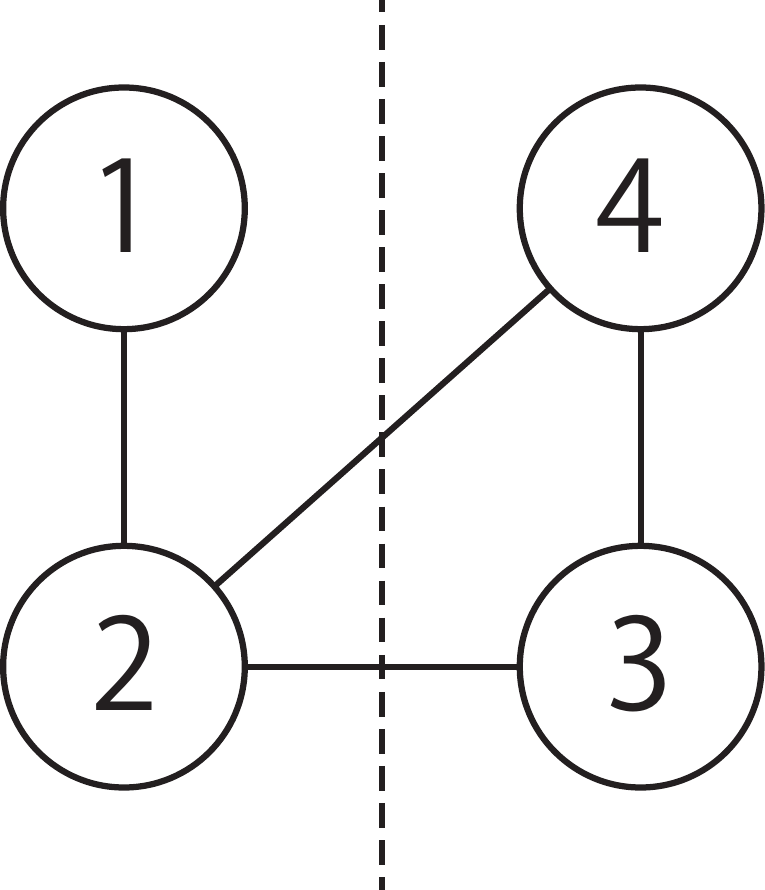}
\caption{
Four-node GPP used in experiments on the gate-based quantum device.
Dashed line denotes the optimum solution.
}
\label{Fig:4vertex_GPP}
\end{figure}
The quantum annealer experiment generates $10$ problem instances for GPPs and $7$ problem instances for QKPs.
For GPPs, $5$ random graphs with $|V_g| \in \{ 32, 64 \}$ and edge density $0.5$ are generated as shown in Sec.~\ref{Sec:Numerical}.
The problem instances are denoted as GPP\_$n$\_$i$ where $n=|V_g|$ and $i$ represents the label of the random graph.
For QKPs, $5$ problems with $50$ items and $2$ problems with $100$ items are generated.
We use the benchmarking instances $100\_100\_i$ to generate the problem instances~\cite{patvardhan2015solving}.
Problem instances with $50$ items are generated in the same process as Sec.~\ref{Sec:Numerical}.
The problem instances are named as QKP\_$n$\_$i$, where $n$ is the number of items and $i$ is the label of the benchmarking instance.
The gate-based quantum device uses a $4$-node GPP (Fig.~\ref{Fig:4vertex_GPP}).
The optimal cost function for each problem is obtained by a heuristic Ising-model classical solver~\cite{fixstars} or reference~\cite{patvardhan2015solving}.

For each problem instance, we generate the Ising model using the QUBOs in~\eqref{Eq:GPP} and~\eqref{Eq:QKP}.
For the experiment on the quantum annealer, auto-scaling of the solver is used to adjust the coefficients of the Ising model.
For the experiments on the gate-based quantum device, the coefficients of the Ising model are rescaled as shown in Sec.~\ref{Sec:Numerical}.

\subsection{Method}
\subsubsection{Implementation on the quantum annealer}~\label{subsubsec:annealer}
\hspace{-1.5mm}We compare the performances of pVSQA, pQA, and QA on the quantum annealer.
Here, VSQA is excluded due to its poor performance compared to pVSQA and pQA (Sec.~\ref{Sec:Numerical}). 
We set the operation time to $5 \mathrm{\mu s}$.
pVSQA adopts the schedule function close to that of the linear schedule.
The schedule function is controlled by a parameter called anneal fraction $s$, which takes a value between $0$ and $1$.
The strength of the transverse fields decreases with $s$.
The values of $s$ at initial time $s_0$ and at end time $s_3$ are fixed to $0$ and $1$, respectively.
We use the values of $s$ at $0.5 \mathrm{\mu s}$ and $4.5 \mathrm{\mu s}$ as variational parameters.
They are denoted by $s_1$ and $s_2$, respectively.
The set of parameters $\{ s_i \}_{i=0}^3$ produces a piecewise linear schedule function.
Two types of classical optimizers are adopted to determine $s_1$ and $s_2$.
One uses the Powell optimizer implemented in SciPy, where the initial values of $s_1$ and $s_2$ are $0.5$ and the maximum number of iterations is $10$.
The other uses the grid-search optimizer to find the optimal values of $s_1$ and $s_2$.
The precision thresholds of $s_1$ and $s_2$ are set to $0.1$.
The probability distribution $p_{50}(\sigma)$ in~\eqref{Eq:modify_probability} is generated by the top $50$ high-quality solutions in terms of the cost function (i.e., $C(\sigma)$) out of the $100$ solutions obtained by the quantum measurements.
pQA and QA use the default schedule function.
Namely, the schedule function begins with $s=0$ and ends with $s=1$ with a linear interpolation between them.

Then we optimize the constraint coefficient $A$ and the chain strength for each problem.
The chain strength is a parameter for minor embedding~\cite{choi2008minor,choi2011minor}.
Here, the parameters are determined as follows.
We obtain $100$ solutions by performing quantum computation along the given schedule function for pQA and QA.
We use the top $50$ high-quality solutions in terms of the cost function (i.e., $C(\sigma)$ for pQA and $C'(\sigma)$ for QA) to compute both the feasible solution rate and the average cost function.
The feasible solution rate is given by the ratio to obtain the feasible solutions in the $50$ solutions.
Then a parameter regime is selected to yield a feasible solution rate above the threshold value of $0.1$.
Afterwards, the optimal values of the constraint coefficient and the chain strength are searched to minimize the average cost function.
The average cost function is given by the average of $C(\sigma)$ and $C'(\sigma)$ over the feasible solutions in the $50$ solutions.
The precision threshold of the constraint coefficient $A$ is $1$ for GPPs and $1000$ for QKPs, whereas that of the chain strength is $10$.
pVSQA uses the same values of $A$ and chain strength as pQA.

Finally, the performance is measured by the success probability and the residual energy.
Success probability $p_{\mathrm{suc}}$ is the ratio to obtain the optimal solutions in the $50$ solutions.
Residual energy $R_{50}^{\mathrm{ave}}$ ($R^{\mathrm{min}}$) is the difference between the average (minimum) cost function and the optimum one.
Smaller values of the residual energies indicate better performances.

In experiments, each algorithm is executed $5$ times for each problem instance.
In Section~\ref{subsec:results}, we show the average and the standard deviation of the success probability and the residual energy based on the $5$ samples.

\subsubsection{Implementation on the gate-based quantum device}~\label{subsubsec:gate}
\hspace{-1.5mm}We compare the performances between pVSQA and VSQA because pQA and QA are difficult to execute on the gate-based quantum device.
The quantum circuit for the QAOA schedule with $p$ layers is given as
\begin{equation}
    \left( \prod_{i=1}^p \hat{U}_{\mathrm{q}}(s_{2i}-s_{2i-1}) \hat{U}_{\mathrm{Ising}}(s_{2i-1}-s_{2i-2}) \right) \ket{\psi(0)}
\end{equation}
where $\hat{U}_{\mathrm{q}}(s) \coloneqq e^{-i \hat{H}_{\mathrm{q}}s}$, $\hat{U}_{\mathrm{Ising}}(s) \coloneqq e^{-i \hat{H}_{\mathrm{Ising}}s}$, and $\ket{\psi(0)}$ is the ground state of $\hat{H}_{\mathrm{q}}$.
The operators are time-ordered (i.e., the operators with $i=1$ are on the right-most, whereas the operators with $i=p$ are on the left-most).
It should be noted that various mixer terms, tailored to explore within the feasible solution subspace~\cite{hadfield2019from}, can be selected as $\hat{U}_{\mathrm{q}}(s)$.
However, for the purpose of this study, the transverse field (i.e., $\hat{U}_{\mathrm{q}}(s) = \exp{(i \sum_{i\in V} \hat{\sigma}_i^x s)}$) is specifically chosen to illustrate the effectiveness of a post-processing approach.
We use Qiskit library to build the quantum circuit~\cite{qiskit2023}.
Here, we adopt the QAOA schedule with $T=1$ and $p=1$.
We set the constraint coefficient $A$ optimized by maximizing the success probability in simulator experiments with the precision threshold of $1$.
We use the Powell optimizer implemented in SciPy as a classical optimizer with a maximum of $10$ iterations.
The initial values of the variational parameters (i.e., $s_1$ and $s_2$) are set to zero.
The probability distribution $p_{100}(\sigma)$ for pVSQA in~\eqref{Eq:modify_probability} and $p_{100}'(\sigma)$ for VQA in~\eqref{Eq:VSQA} are generated by the $50$ solutions obtained by the quantum measurements.
Finally, the performance on the gate-based quantum device is measured by the success probability from $50$ solutions.

In experiments, each algorithm is executed $3$ times.
Below, we show the average and the standard deviation of the success probability based on the $3$ samples.

\subsection{Results}~\label{subsec:results}
\hspace{-1.5mm}Table~\ref{Table:DWave} displays the performances of pVSQA, pQA, and QA on the quantum annealer.
Post-processing significantly improves the performance.
The choice of the classical optimizer influences the pVSQA performance.
The grid-search optimizer outperforms the Powell optimizer.
pVSQA with the grid-search optimizer yields better results than pQA for all problem instances, but pVSQA with the Powell optimizer exhibits performance similar to pQA, particularly for QKPs.
pVSQA with the grid-search optimizer finds optimal solutions, except for GPPs with $64$ nodes.
Even for problems with low (or no) success probability, pVSQA with the grid-search optimizer produces smaller residual energies ($R_{50}^{\mathrm{ave}}$ and $R^{\mathrm{min}}$) than pQA and QA.

Table~\ref{Table:IBMQ} shows the performances of pVSQA and VSQA on the gate-based quantum device.
pVSQA outperforms VSQA.
The success probability of pVSQA on the quantum device is slightly worse than that on the simulator.
This difference is attributed to the inherent noise in the quantum device.

\begin{table*}[ht]
  \centering 
  \caption{Experimental results of pVSQA, pQA, and QA for GPPs and QKPs on the quantum annealer.
  Average residual energy, minimum residual energy, success probability, and feasible solution rate are denoted by $R_{50}^{\mathrm{ave}}$, $R^{\mathrm{min}}$, $p_{\mathrm{suc}}$, and $p_{\mathrm{FS}}$, respectively.
  Values in parentheses denote the standard deviation of the average.
  Average indicates the average of $R_{50}^{\mathrm{ave}}$, $R^{\mathrm{min}}$, and $p_{\mathrm{suc}}$ over the GPPs and QKPs with the same problem size.
  Opt denotes the cost function of the optimal solution.
   } 
  \scalebox{1}{
  \begin{tabular}{c|c|ccccc|ccccc}\hline
    \hline
   \multicolumn{1}{c|}{} & \multicolumn{1}{c|}{} & \multicolumn{5}{c|}{pVSQA (Powell)} & \multicolumn{5}{c}{pVSQA (grid search)} \\ \hline
   instance   & Opt & $R_{\mathrm{50}}^{\mathrm{ave}}$ & $R^{\mathrm{min}}$ & $p_{\mathrm{suc}}$ & $s_1$ & $s_2$ & $R_{\mathrm{50}}^{\mathrm{ave}}$ & $R^{\mathrm{min}}$ & $p_{\mathrm{suc}}$ & $s_1$ & $s_2$ \\ \hline\hline
   GPP\_32\_1 & $91$  & $0.19 (8)$ & $0.0 (0)$ & $0.82 (7)$ & $0.32(9)$ & $0.59(6)$ & $0.02(2)$ & $0.0 (0)$ & $0.98 (2)$ & $0.3$ & $0.7$ \\
   GPP\_32\_2 & $92$  & $2.62 (29)$ & $0.0 (0)$ & $0.11 (2)$ & $0.38 (13)$ & $0.65 (8)$ & $2.12 (9)$ & $0.0 (0)$ & $0.13 (2)$ & $0.0$ & $0.6$ \\
   GPP\_32\_3 & $93$  & $4.58 (25)$ & $1.2 (4)$ & $0.01 (1)$ & $0.37 (10)$ & $0.63 (8)$ & $4.14 (1)$ & $1.2 (5)$ & $0.01 (1)$ & $0$ & $0.5$ \\
   GPP\_32\_4 & $89$  & $2.32 (38)$ & $0.0 (0)$ & $0.14 (2)$ & $0.40 (9)$ & $0.59 (5)$ & $1.68 (11)$ & $0.0 (0)$ & $0.22 (2)$ & $0.2$ & $0.5$ \\
   GPP\_32\_5 & $104$  & $2.18 (48)$ & $0.0 (0)$ & $0.18 (4)$ & $0.59 (8)$ & $0.57 (8)$ & $1.66 (15)$ & $0.0 (0)$ & $0.30 (2)$ & $0.2$ & $1.0$ \\ \hline
   Average & $-$  & $2.38$ & $0.2$ & $0.25$ & $-$ & $-$ & $1.92$ & $0.2$ & $0.33$ & $-$ & $-$ \\ \hline \hline
   GPP\_64\_1 & $428$  & $6.44 (41)$ & $4.0 (6)$ & $0.0 (0)$ & $0.63 (12)$ & $0.44 (8)$ & $5.87 (8)$ & $4.0 (5)$ & $0.0 (0)$ & $0.3$ & $0.3$ \\
   GPP\_64\_2 & $437$  & $9.59 (21)$ & $3.4 (7)$ & $0.0 (0)$ & $0.54 (7)$ & $0.54 (4)$ & $8.18 (52)$ & $3.6 (7)$& $0.0 (0)$ & $1.0$ & $0.0$ \\
   GPP\_64\_3 & $425$  & $12.5 (3)$ & $6.8 (17)$ & $0.004 (4)$ & $0.42 (11)$ & $0.55 (12)$ & $11.7 (2)$ & $8.0 (3)$ & $0.0 (0)$ & $0.9$ & $0.0$ \\
   GPP\_64\_4 & $407$  & $12.9 (23)$ & $6.0 (13)$ & $0.0 (0)$ & $0.51 (8)$ & $0.33 (10)$ & $10.2 (3)$ & $4.2 (5)$ & $0.0 (0)$ & $0.6$ & $0.1$ \\
   GPP\_64\_5 & $431$  & $10.4 (3)$ & $3.2 (7)$ & $0.0 (0)$ & $0.62 (5)$ & $0.57 (9)$ & $9.01 (49)$ & $1.8 (5)$& $0.004 (4)$ & $0.9$ & $0.0$ \\ \hline
   Average & $-$  & $10.4$ & $4.7$ & $0.001$ & $-$ & $-$ & $8.99$ & $4.3$ & $0.001$ & $-$ & $-$ \\ \hline \hline
   QKP\_50\_1 & $-21790$  & $485 (81)$ & $274 (78)$ & $0.004 (4)$ & $0.65 (5)$ & $0.58 (11)$ & $307 (21)$ & $199 (72)$ & $0.04 (4)$ & $0.6$ & $0.4$ \\
   QKP\_50\_2 & $-47302$  & $169 (8)$ & $0 (0)$ & $0.10 (3)$ & $0.37 (10)$ & $0.45 (10)$ & $108 (21)$ & $0 (0)$ & $0.36 (7)$ & $0.3$ & $0.6$ \\
   QKP\_50\_3 & $-58111$  & $173 (95)$ & $21 (21)$ & $0.48 (22)$ & $0.67 (4)$ & $0.43 (8)$ & $0 (0)$ & $0 (0)$ & $1.0 (0)$ & $0.6$ & $1.0$ \\
   QKP\_50\_4 & $-17991$  & $285 (46)$ & $33 (20)$ & $0.02 (1)$ & $0.42 (8)$ & $0.36 (2)$ & $183 (34)$ & $46 (29)$ & $0.16 (11)$ & $0.5$ & $0.3$ \\
   QKP\_50\_5 & $-49741$  & $214 (64)$ & $127 (0)$ & $0.0 (0)$ & $0.82 (4)$ & $0.64 (10)$ & $112 (14)$ & $76 (31)$ & $0.008 (5)$ & $0.5$ & $0.2$ \\ \hline
   Average & $-$  & $265$ & $91$ & $0.12$ & $-$ & $-$ & $142$ & $64$ & $0.31$ & $-$ & $-$ \\ \hline \hline
   QKP\_100\_1 & $-81978$  & $114 (28)$ & $17 (0)$ & $0.0 (0)$ & $0.52 (10)$ & $0.49 (10)$ & $29 (6)$ & $10 (4)$ & $0.06 (5)$ & $0.7$ & $0.4$ \\
   QKP\_100\_2 & $-190424$  & $312 (23)$ & $89 (31)$ & $0.004 (4)$ & $0.54 (5)$ & $0.39 (7)$ & $218 (27)$ & $80 (32)$ & $0.004 (4)$ & $0.2$ & $0.5$ \\ \hline
   Average & $-$  & $213$ & $53$ & $0.002$ & $-$ & $-$ & $124$ & $45$ & $0.03$ & $-$ & $-$ \\ \hline \hline
   \cline{1-9}
   \multicolumn{1}{c|}{} & \multicolumn{1}{c|}{} & \multicolumn{3}{c|}{pQA} & \multicolumn{4}{c}{QA} & \multicolumn{3}{c}{} \\ \cline{1-9}
   \multicolumn{1}{c|}{instance} & \multicolumn{1}{c|}{Opt} & \multicolumn{1}{c}{$R_{\mathrm{50}}^{\mathrm{ave}}$} & \multicolumn{1}{c}{$R^{\mathrm{min}}$} & \multicolumn{1}{c|}{$p_{\mathrm{suc}}$} & \multicolumn{1}{c}{$R_{\mathrm{50}}^{\mathrm{ave}}$} & \multicolumn{1}{c}{$R^{\mathrm{min}}$} & \multicolumn{1}{c}{$p_{\mathrm{FS}}$} & \multicolumn{1}{c}{$p_{\mathrm{suc}}$} & \multicolumn{3}{c}{} \\ \cline{1-9}
   \multicolumn{1}{c|}{GPP\_32\_1} & \multicolumn{1}{c|}{$91$} & \multicolumn{1}{c}{$0.38 (8)$} & \multicolumn{1}{c}{$0.0 (0)$} & \multicolumn{1}{c|}{$0.66 (6)$} & \multicolumn{1}{c}{$15.6 (7)$} & \multicolumn{1}{c}{$10.0 (9)$} & \multicolumn{1}{c}{$0.27 (2)$} & \multicolumn{1}{c}{$0.0(0)$} & \multicolumn{3}{c}{} \\
   \multicolumn{1}{c|}{GPP\_32\_2} & \multicolumn{1}{c|}{$92$} & \multicolumn{1}{c}{$3.20 (30)$} & \multicolumn{1}{c}{$0.0 (0)$} & \multicolumn{1}{c|}{$0.08 (1)$} & \multicolumn{1}{c}{$20.9 (7)$} & \multicolumn{1}{c}{$14.4 (20)$} & \multicolumn{1}{c}{$0.22 (4)$} & \multicolumn{1}{c}{$0.0(0)$} & \multicolumn{3}{c}{} \\
   \multicolumn{1}{c|}{GPP\_32\_3} & \multicolumn{1}{c|}{$93$} & \multicolumn{1}{c}{$4.40 (20)$} & \multicolumn{1}{c}{$1.4 (4)$} & \multicolumn{1}{c|}{$0.004 (4)$} & \multicolumn{1}{c}{$19.3 (6)$} & \multicolumn{1}{c}{$14.0 (9)$} & \multicolumn{1}{c}{$0.40 (1)$} & \multicolumn{1}{c}{$0.0(0)$} & \multicolumn{3}{c}{} \\
   \multicolumn{1}{c|}{GPP\_32\_4} & \multicolumn{1}{c|}{$89$} & \multicolumn{1}{c}{$2.54 (19)$} & \multicolumn{1}{c}{$0.0 (0)$} & \multicolumn{1}{c|}{$0.12 (3)$} & \multicolumn{1}{c}{$17.8 (3)$} & \multicolumn{1}{c}{$11.4 (14)$} & \multicolumn{1}{c}{$0.23 (2)$} & \multicolumn{1}{c}{$0.0(0)$} & \multicolumn{3}{c}{} \\
   \multicolumn{1}{c|}{GPP\_32\_5} & \multicolumn{1}{c|}{$104$} & \multicolumn{1}{c}{$2.16 (16)$} & \multicolumn{1}{c}{$0.0 (0)$} & \multicolumn{1}{c|}{$0.22 (2)$} & \multicolumn{1}{c}{$17.3 (4)$} & \multicolumn{1}{c}{$8.8 (11)$} & \multicolumn{1}{c}{$0.22 (3)$} & \multicolumn{1}{c}{$0.0(0)$} & \multicolumn{3}{c}{} \\ \cline{1-9}
   \multicolumn{1}{c|}{Average} & \multicolumn{1}{c|}{$-$} & \multicolumn{1}{c}{$2.54$} & \multicolumn{1}{c}{$0.3$} & \multicolumn{1}{c|}{$0.22$} & \multicolumn{1}{c}{$18.2$} & \multicolumn{1}{c}{$11.7$} & \multicolumn{1}{c}{$0.27$} & \multicolumn{1}{c}{$0.0$} & \multicolumn{3}{c}{} \\ \cline{1-9} \cline{1-9}
   \multicolumn{1}{c|}{GPP\_64\_1} & \multicolumn{1}{c|}{$428$} & \multicolumn{1}{c}{$6.00 (0)$} & \multicolumn{1}{c}{$6.0 (0)$} & \multicolumn{1}{c|}{$0.0 (0)$} & \multicolumn{1}{c}{$56.5 (12)$} & \multicolumn{1}{c}{$49.6 (17)$} & \multicolumn{1}{c}{$0.14 (2)$} & \multicolumn{1}{c}{$0.0(0)$} & \multicolumn{3}{c}{} \\
   \multicolumn{1}{c|}{GPP\_64\_2} & \multicolumn{1}{c|}{$437$} & \multicolumn{1}{c}{$11.1 (2)$} & \multicolumn{1}{c}{$4.4 (6)$} & \multicolumn{1}{c|}{$0.0 (0)$} & \multicolumn{1}{c}{$61.7 (16)$} & \multicolumn{1}{c}{$51.2 (44)$} & \multicolumn{1}{c}{$0.17 (4)$} & \multicolumn{1}{c}{$0.0(0)$} & \multicolumn{3}{c}{} \\
   \multicolumn{1}{c|}{GPP\_64\_3} & \multicolumn{1}{c|}{$425$} & \multicolumn{1}{c}{$14.4(1)$} & \multicolumn{1}{c}{$9.2 (2)$} & \multicolumn{1}{c|}{$0.0 (0)$} & \multicolumn{1}{c}{$56.1 (9)$} & \multicolumn{1}{c}{$45.6 (17)$} & \multicolumn{1}{c}{$0.21 (3)$} & \multicolumn{1}{c}{$0.0(0)$} & \multicolumn{3}{c}{} \\
   \multicolumn{1}{c|}{GPP\_64\_4} & \multicolumn{1}{c|}{$407$} & \multicolumn{1}{c}{$11.4 (1)$} & \multicolumn{1}{c}{$6.0 (8)$} & \multicolumn{1}{c|}{$0.0 (0)$} & \multicolumn{1}{c}{$57.5 (10)$} & \multicolumn{1}{c}{$47.4 (20)$} & \multicolumn{1}{c}{$0.20 (4)$} & \multicolumn{1}{c}{$0.0(0)$} & \multicolumn{3}{c}{} \\
   \multicolumn{1}{c|}{GPP\_64\_5} & \multicolumn{1}{c|}{$431$} & \multicolumn{1}{c}{$11.0 (7)$} & \multicolumn{1}{c}{$6.6 (21)$} & \multicolumn{1}{c|}{$0.0 (0)$} & \multicolumn{1}{c}{$53.3 (12)$} & \multicolumn{1}{c}{$42.0 (31)$} & \multicolumn{1}{c}{$0.18 (3)$} & \multicolumn{1}{c}{$0.0(0)$} & \multicolumn{3}{c}{} \\ \cline{1-9}
   \multicolumn{1}{c|}{Average} & \multicolumn{1}{c|}{$-$} & \multicolumn{1}{c}{$10.8$} & \multicolumn{1}{c}{$6.4$} & \multicolumn{1}{c|}{$0.0$} & \multicolumn{1}{c}{$57.0$} & \multicolumn{1}{c}{$47.2$} & \multicolumn{1}{c}{$0.18$} & \multicolumn{1}{c}{$0.0$} & \multicolumn{3}{c}{} \\ \cline{1-9}\cline{1-9}
   \multicolumn{1}{c|}{QKP\_50\_1} & \multicolumn{1}{c|}{$-21790$} & \multicolumn{1}{c}{$331 (23)$} & \multicolumn{1}{c}{$203(57)$} & \multicolumn{1}{c|}{$0.0 (0)$} & \multicolumn{1}{c}{$497 (180)$} & \multicolumn{1}{c}{$117 (81)$} & \multicolumn{1}{c}{$0.43 (12)$} & \multicolumn{1}{c}{$0.04(3)$} & \multicolumn{3}{c}{} \\
   \multicolumn{1}{c|}{QKP\_50\_2} & \multicolumn{1}{c|}{$-47302$} & \multicolumn{1}{c}{$164 (20)$} & \multicolumn{1}{c}{$0(0)$} & \multicolumn{1}{c|}{$0.18 (7)$} & \multicolumn{1}{c}{$1420 (339)$} & \multicolumn{1}{c}{$841 (118)$} & \multicolumn{1}{c}{$0.81 (10)$} & \multicolumn{1}{c}{$0.0(0)$} & \multicolumn{3}{c}{} \\
   \multicolumn{1}{c|}{QKP\_50\_3} & \multicolumn{1}{c|}{$-58111$} & \multicolumn{1}{c}{$37 (23)$} & \multicolumn{1}{c}{$19 (19)$} & \multicolumn{1}{c|}{$0.61 (24)$} & \multicolumn{1}{c}{$2599 (307)$} & \multicolumn{1}{c}{$1645 (614)$} & \multicolumn{1}{c}{$0.99 (1)$} & \multicolumn{1}{c}{$0.0(0)$} & \multicolumn{3}{c}{} \\
   \multicolumn{1}{c|}{QKP\_50\_4} & \multicolumn{1}{c|}{$-17991$} & \multicolumn{1}{c}{$272 (7)$} & \multicolumn{1}{c}{$93 (54)$} & \multicolumn{1}{c|}{$0.008 (4)$} & \multicolumn{1}{c}{$2757 (574)$} & \multicolumn{1}{c}{$1463 (518)$} & \multicolumn{1}{c}{$0.65 (13)$} & \multicolumn{1}{c}{$0.0(0)$} & \multicolumn{3}{c}{} \\
   \multicolumn{1}{c|}{QKP\_50\_5} & \multicolumn{1}{c|}{$-49741$} & \multicolumn{1}{c}{$179 (52)$} & \multicolumn{1}{c}{$127 (0)$} & \multicolumn{1}{c|}{$0.0 (0)$} & \multicolumn{1}{c}{$1278 (409)$} & \multicolumn{1}{c}{$239 (68)$} & \multicolumn{1}{c}{$0.93 (3)$} & \multicolumn{1}{c}{$0.0(0)$} & \multicolumn{3}{c}{} \\ \cline{1-9}
   \multicolumn{1}{c|}{Average} & \multicolumn{1}{c|}{$-$} & \multicolumn{1}{c}{$197$} & \multicolumn{1}{c}{$88$} & \multicolumn{1}{c|}{$0.16$} & \multicolumn{1}{c}{$1710$} & \multicolumn{1}{c}{$861$} & \multicolumn{1}{c}{$0.76$} & \multicolumn{1}{c}{$0.01$} & \multicolumn{3}{c}{} \\ \cline{1-9} \cline{1-9}
   \multicolumn{1}{c|}{QKP\_100\_1} & \multicolumn{1}{c|}{$-81978$} & \multicolumn{1}{c}{$63 (27)$} & \multicolumn{1}{c}{$17 (0)$} & \multicolumn{1}{c|}{$0.0 (0)$} & \multicolumn{1}{c}{$5286 (1393)$} & \multicolumn{1}{c}{$2576 (1053)$} & \multicolumn{1}{c}{$0.47 (13)$} & \multicolumn{1}{c}{$0.0(0)$} & \multicolumn{3}{c}{} \\
   \multicolumn{1}{c|}{QKP\_100\_2} & \multicolumn{1}{c|}{$-190424$} & \multicolumn{1}{c}{$317 (49)$} & \multicolumn{1}{c}{$110 (28)$} & \multicolumn{1}{c|}{$0.0 (0)$} & \multicolumn{1}{c}{$1089 (216)$} & \multicolumn{1}{c}{$283 (98)$} & \multicolumn{1}{c}{$0.45 (12)$} & \multicolumn{1}{c}{$0.0(0)$} & \multicolumn{3}{c}{} \\ \cline{1-9}
   \multicolumn{1}{c|}{Average} & \multicolumn{1}{c|}{$-$} & \multicolumn{1}{c}{$190$} & \multicolumn{1}{c}{$64$} & \multicolumn{1}{c|}{$0.0$} & \multicolumn{1}{c}{$3188$} & \multicolumn{1}{c}{$1430$} & \multicolumn{1}{c}{$0.46$} & \multicolumn{1}{c}{$0.0$} & \multicolumn{3}{c}{} \\ \cline{1-9} \cline{1-9}
  \end{tabular}
  }
  \label{Table:DWave}
\end{table*}

\begin{figure*}[th]
\centering
\begin{tabular}{p{0.17\linewidth}p{0.17\linewidth}p{0.17\linewidth}p{0.17\linewidth}p{0.17\linewidth}}
	\begin{minipage}[b]{\linewidth}
		\centering
		\includegraphics[width=\linewidth]{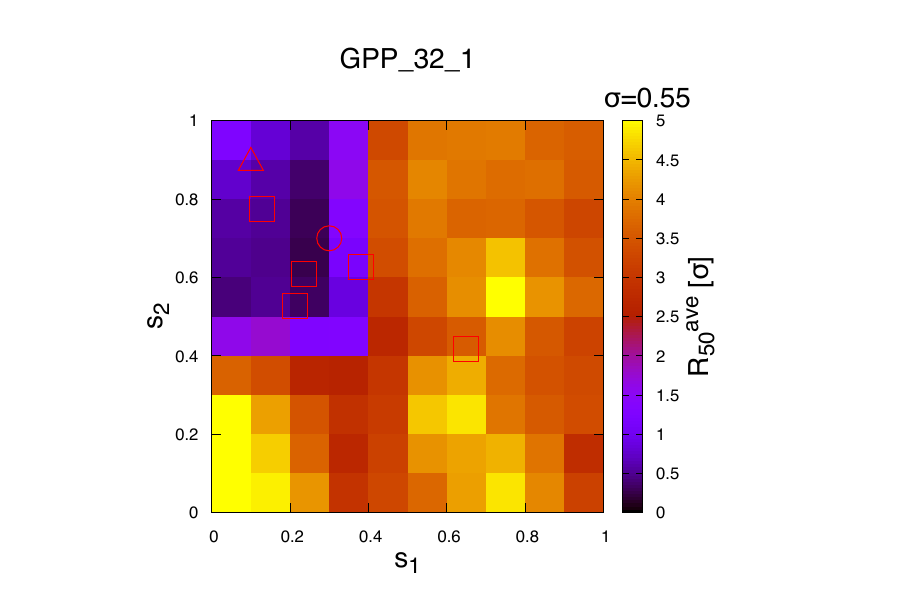}
		\end{minipage} &
	\begin{minipage}[b]{\linewidth}
		\centering
		\includegraphics[width=\linewidth]{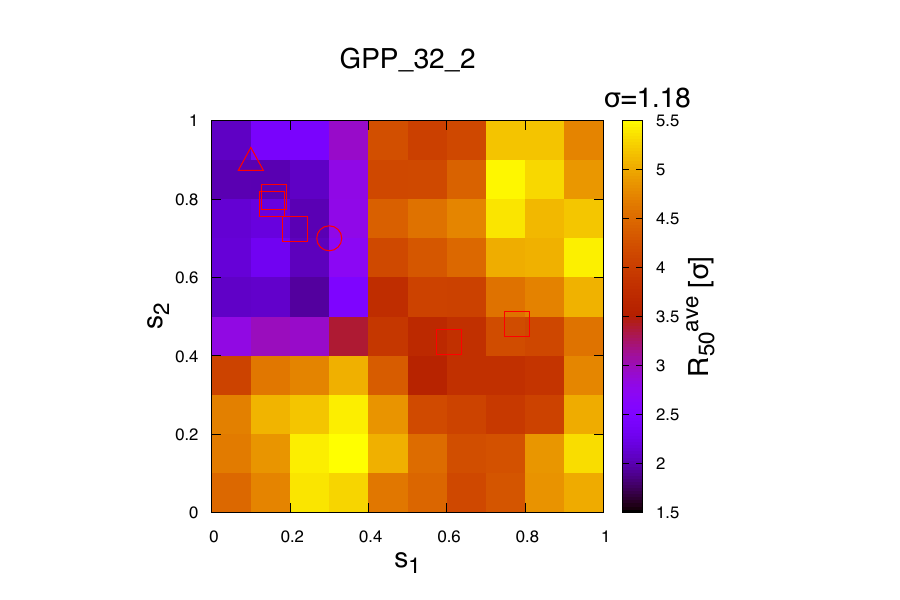}
		\end{minipage} &
	\begin{minipage}[b]{\linewidth}
		\centering
		\includegraphics[width=\linewidth]{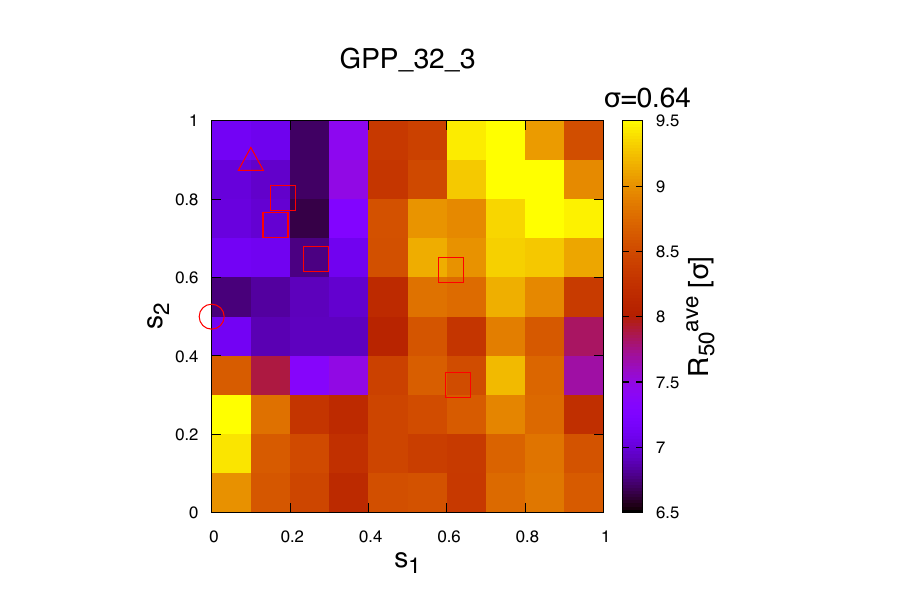}
		\end{minipage} &
        \begin{minipage}[b]{\linewidth}
		\centering
		\includegraphics[width=\linewidth]{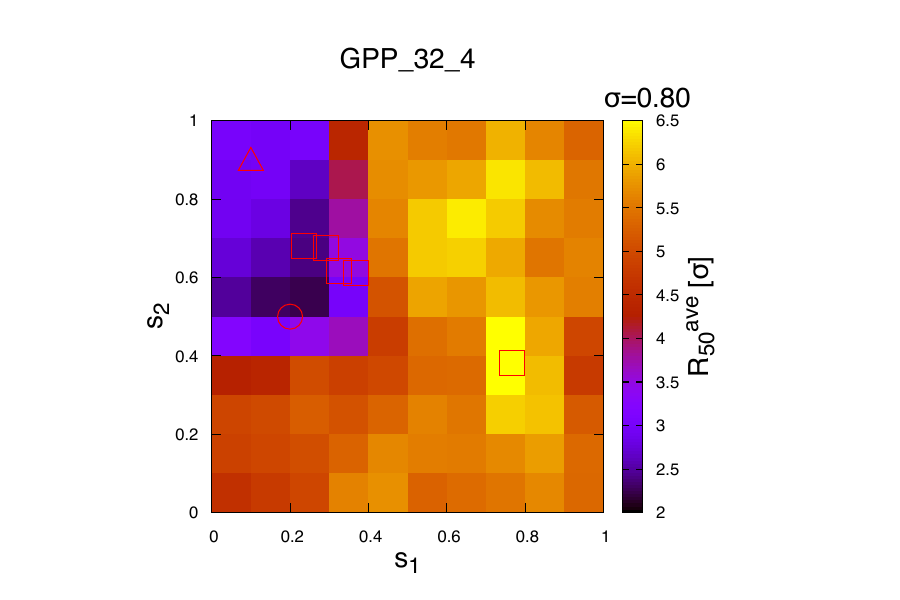}
		\end{minipage} &
	\begin{minipage}[b]{\linewidth}
		\centering
		\includegraphics[width=\linewidth]{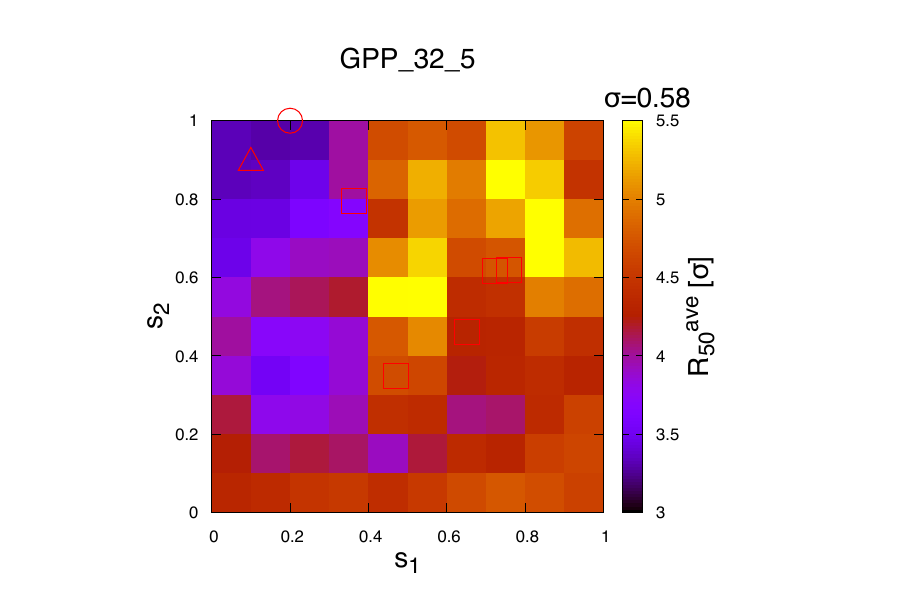}
		\end{minipage}
	\\ \cline{1-5}\\
	    \begin{minipage}[b]{\linewidth}
		\centering
		\includegraphics[width=\linewidth]{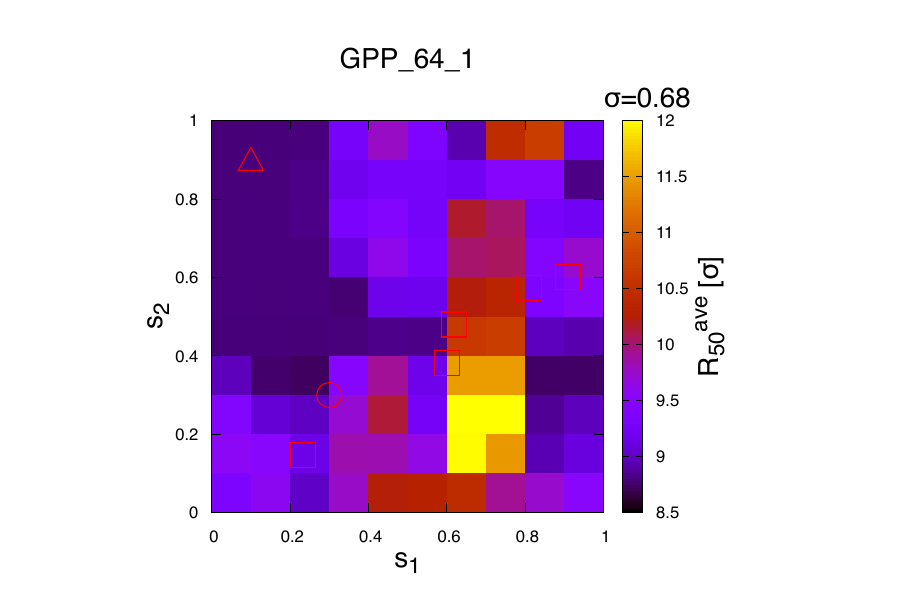}
		\end{minipage} &
	\begin{minipage}[b]{\linewidth}
		\centering
		\includegraphics[width=\linewidth]{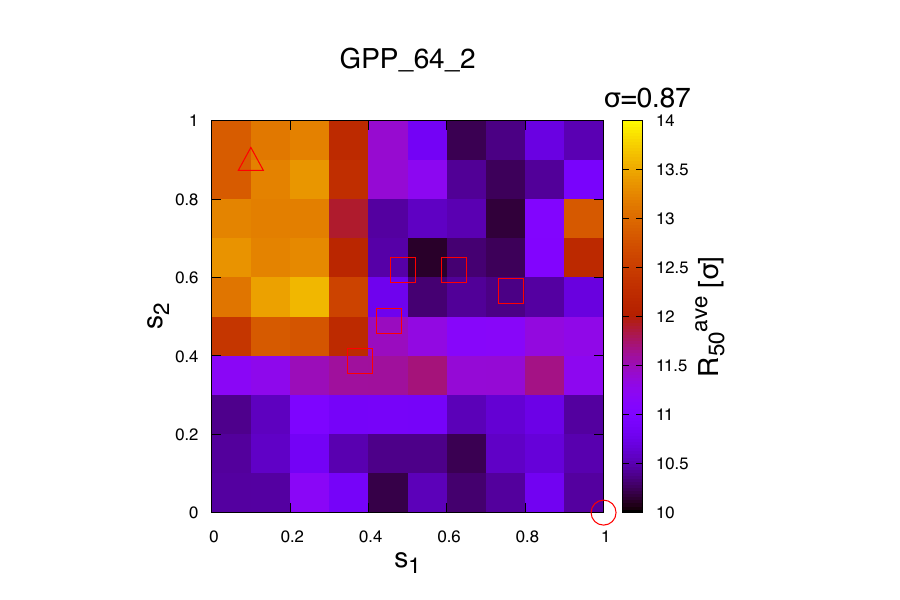}
		\end{minipage} &
	\begin{minipage}[b]{\linewidth}
		\centering
		\includegraphics[width=\linewidth]{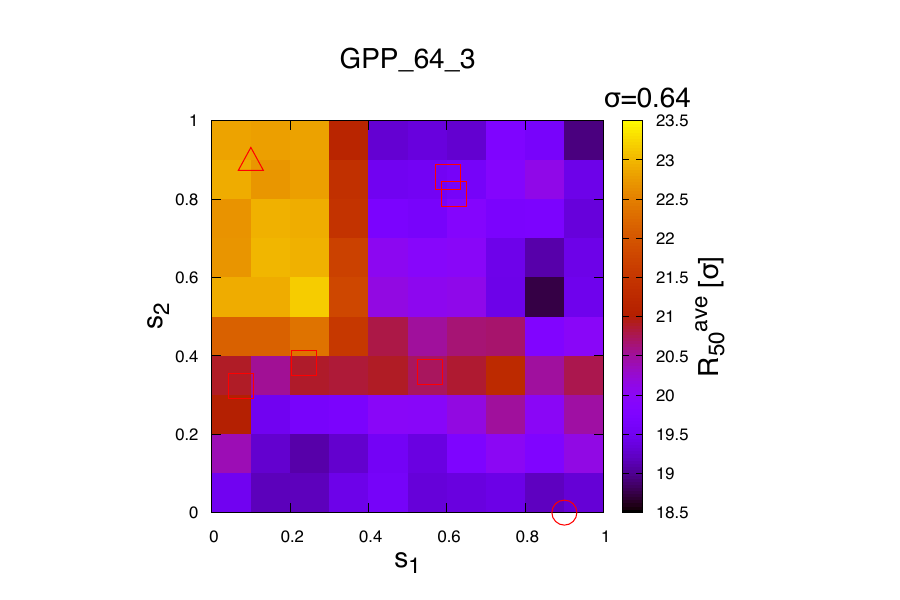}
		\end{minipage} &
        \begin{minipage}[b]{\linewidth}
		\centering
		\includegraphics[width=\linewidth]{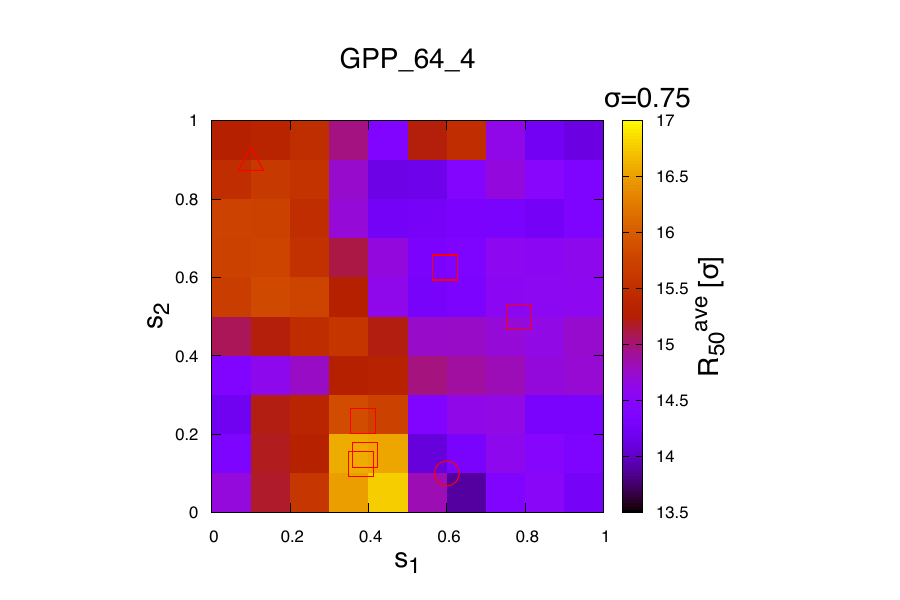}
		\end{minipage} &
	\begin{minipage}[b]{\linewidth}
		\centering
		\includegraphics[width=\linewidth]{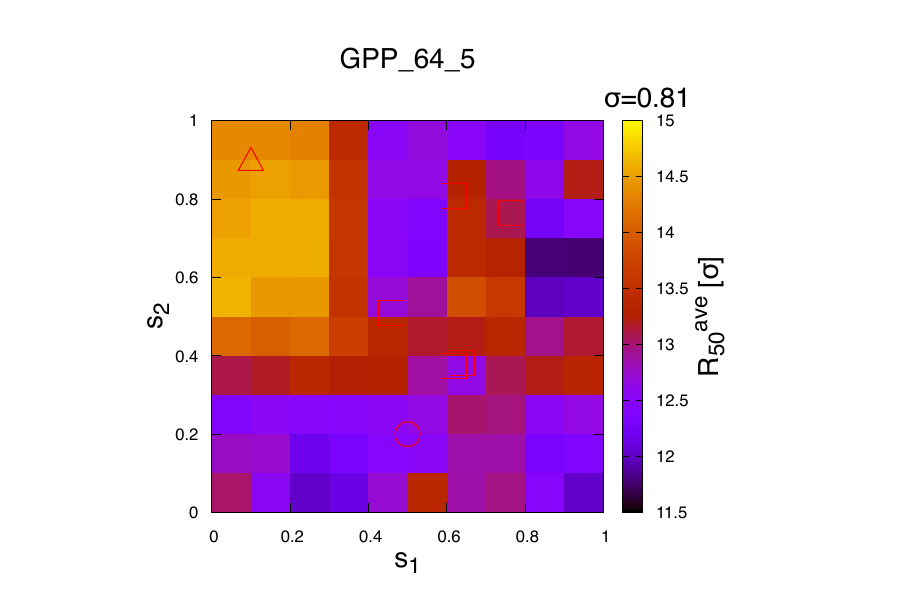}
		\end{minipage}
	\\ \cline{1-5}\\
	\begin{minipage}[b]{\linewidth}
		\centering
		\includegraphics[width=\linewidth]{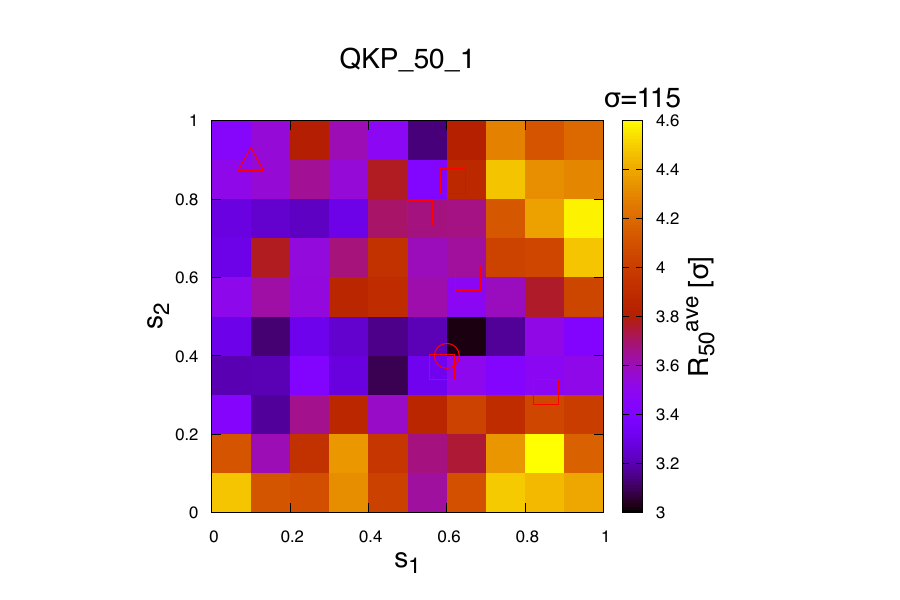}
		\end{minipage} &
	\begin{minipage}[b]{\linewidth}
		\centering
		\includegraphics[width=\linewidth]{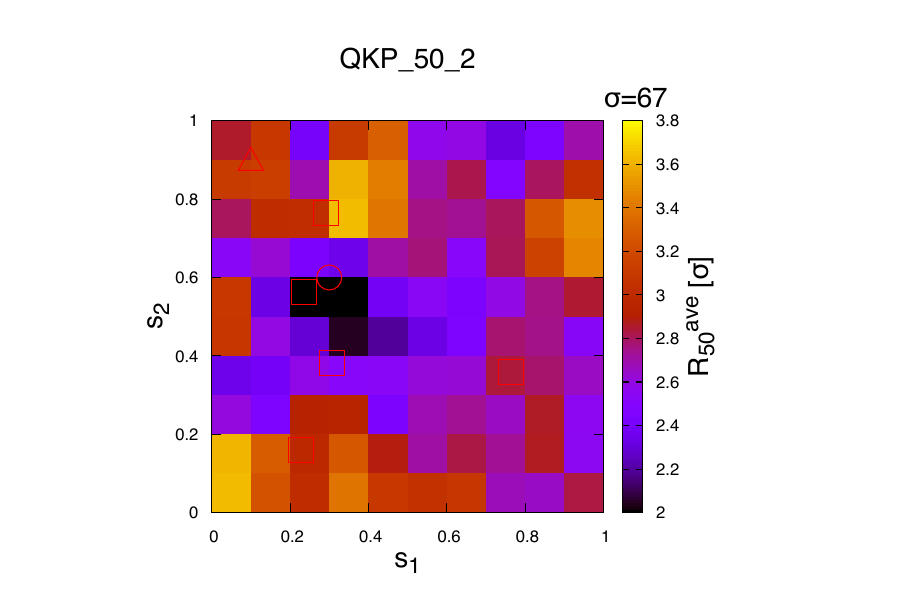}
		\end{minipage} &
	\begin{minipage}[b]{\linewidth}
		\centering
		\includegraphics[width=\linewidth]{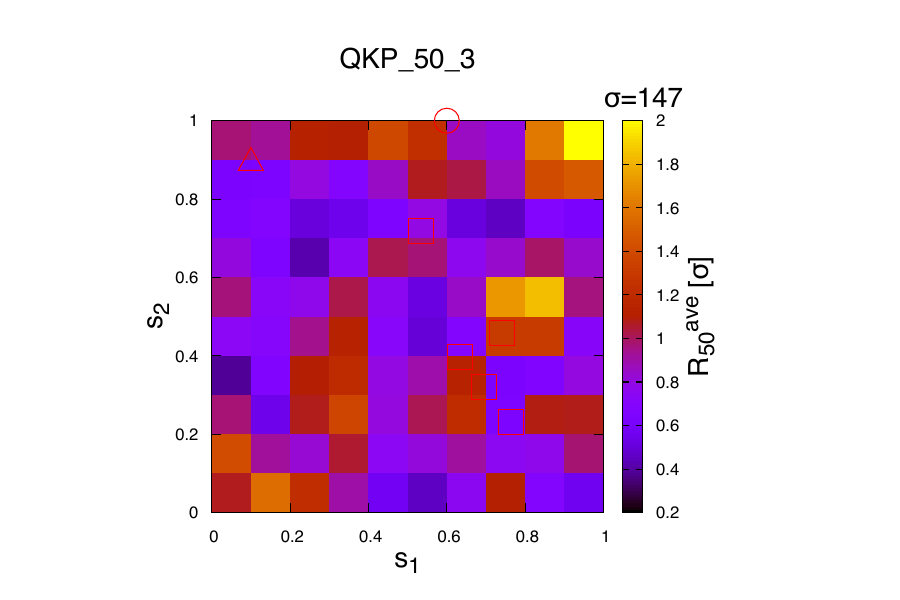}
		\end{minipage} &
        \begin{minipage}[b]{\linewidth}
		\centering
		\includegraphics[width=\linewidth]{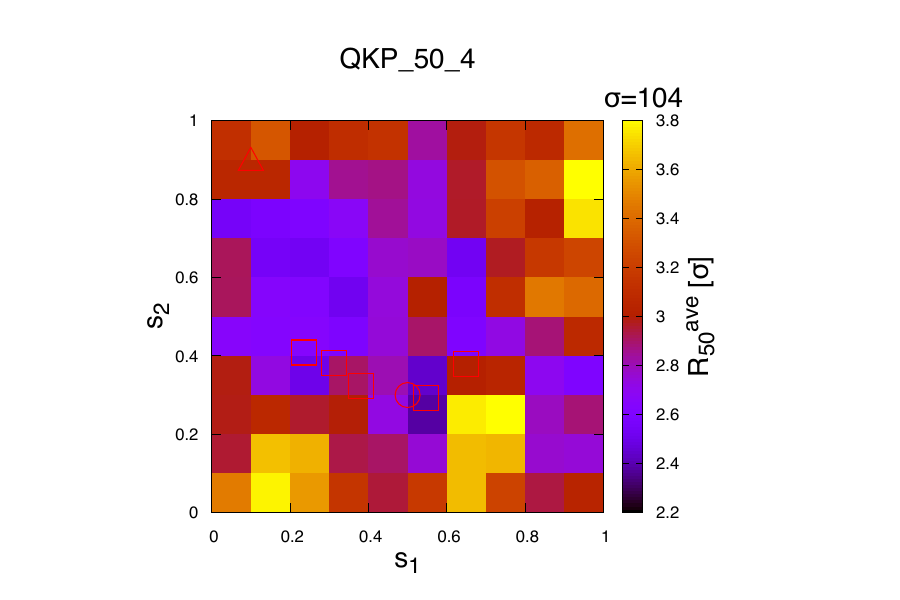}
		\end{minipage} &
	\begin{minipage}[b]{\linewidth}
		\centering
		\includegraphics[width=\linewidth]{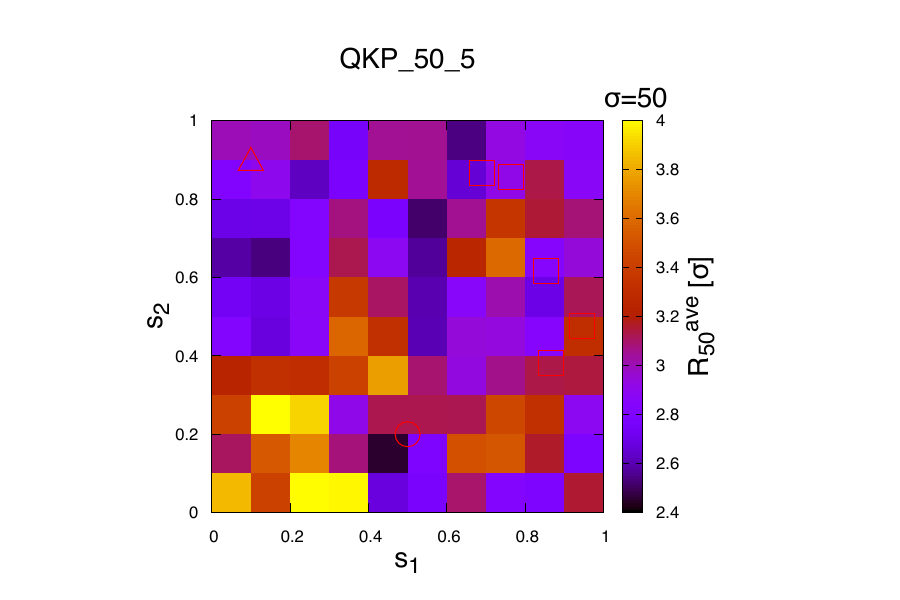}
		\end{minipage}
 \\ \cline{1-5}\\
	\begin{minipage}[b]{\linewidth}
		\centering
		\includegraphics[width=\linewidth]{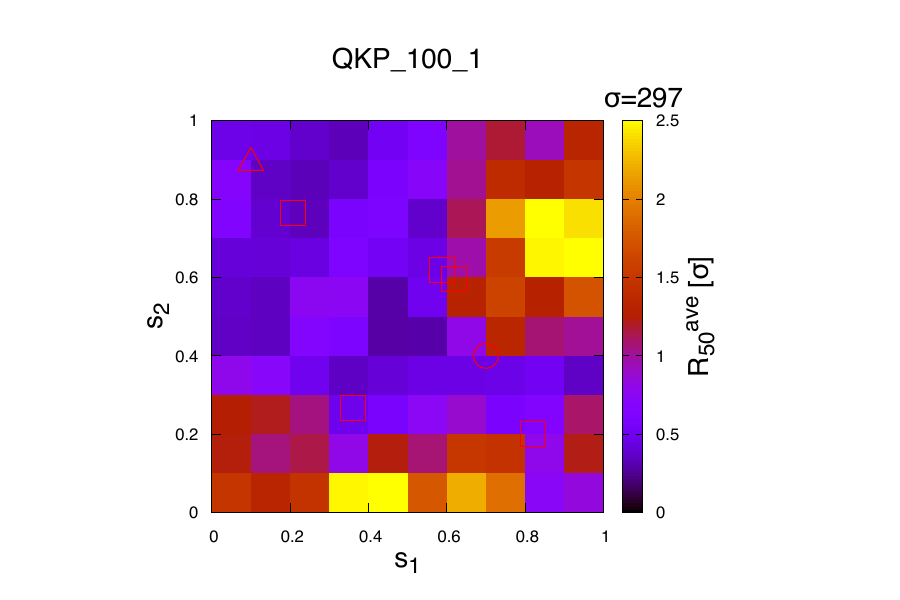}
		\end{minipage} &
	\begin{minipage}[b]{\linewidth}
		\centering
		\includegraphics[width=\linewidth]{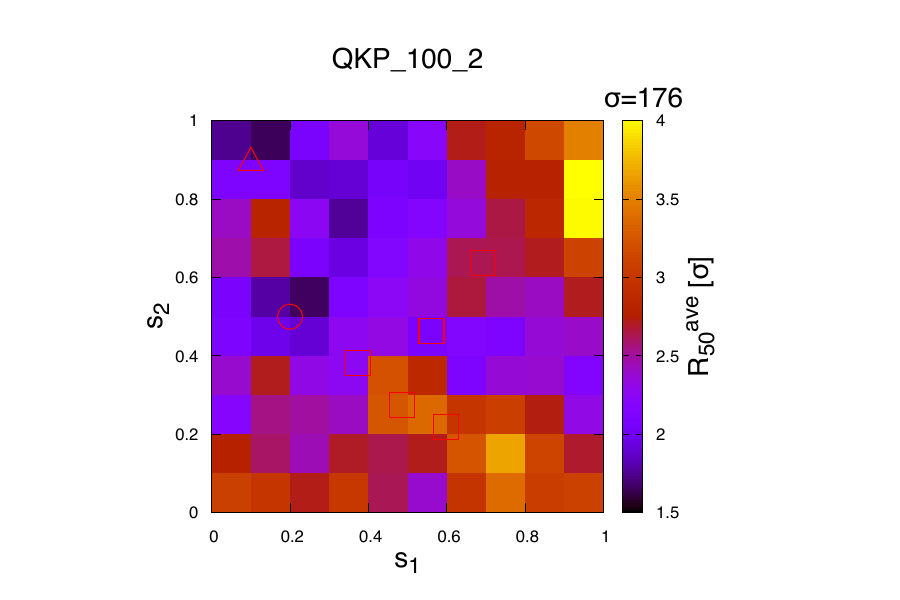}
		\end{minipage}
\end{tabular}
\caption{Dependences of $R_{50}^{\mathrm{ave}}$ on $s_1$ and $s_2$ for GPPs and QKPs. The color in each mesh represents the mean of $R_{50}^{\mathrm{ave}}$ at the surrounding corners.
Circles and squares denote the optimal values of $s_1$ and $s_2$ obtained using the grid-search optimizer and the Powell optimizer, respectively.
Triangles correspond to $s_1$ and $s_2$ values in pQA.}
\label{fig:sdep}
\end{figure*}

\begin{table}[ht]
  \centering 
  \caption{Success probability of pVSQA and VSQA for the GPP on the gate-based quantum device (qpu) and the simulator (cpu).
  Values in parentheses denote the standard deviation of the average.
   } 
  \scalebox{1}{
  \begin{tabular}{c|c|c|c}\hline
  pVSQA (qpu) & VSQA (qpu) & pVSQA (cpu) & VSQA (cpu)\\ \hline \hline
      $0.91 (4)$ & $0.15 (3)$ & $1.00 (0)$ & $0.35 (6)$ \\
    \hline
  \end{tabular}
  }
  \label{Table:IBMQ}
\end{table}

\section{Discussion}~\label{Sec:Discussion}
\hspace{-1.5mm}The simulator result shows that the linear schedule almost reproduces the success probability of the continuous schedule in both GPPs and QKPs.
This is expected with short or long operation times.
Because the quantum state does not change significantly from the initial one when the operation time is short, the success probability is close to the random sampling with a post-processing.
This situation is almost independent of the schedule function.
When the operation time is long, the conventional quantum annealing schedule (i.e., linear schedule with $s_0=0$ and $s_1=1$) achieves the ground state.
Consequently, a small modification of the schedule function does not affect the performance.
However, this result is nontrivial for an intermediate operation time.
Although the optimum schedule function for the continuous schedule is a complicated bang-anneal-bang type (see Fig.~\ref{Fig:schedule} (b)), a simple linear schedule gives a good approximation.
The QAOA schedule also provides a suitable approximation of the continuous schedule in the bang-bang regime and the bang-anneal-bang regime.
These results indicate that a small number of variational parameters is sufficient to implement pVSQA.

The experimental results on a quantum annealer demonstrate the effectiveness of our proposed algorithm in both GPP and QKP.
To investigate these results in more detail, we illustrate the grid-search results of $R_{50}^{\mathrm{ave}}$ for both GPP and QKP in Fig.~\ref{fig:sdep}.
They are rescaled in units of their respective $\sigma$ values, the standard deviations of $R_{50}^{\mathrm{ave}}$ averaged over $s_1$ and $s_2$.
Circles and squares represent the optimal values of $s_1$ and $s_2$ obtained using the grid-search optimizer and the Powell optimizer, respectively.
Triangles correspond to $s_1$ and $s_2$ values in pQA (Note that the default schedule function is used in pQA experiments and $s_1$ and $s_2$ are not explicitly specified).
For GPPs, the upper-left regime (i.e., small $s_1$ and large $s_2$) yields lower energies for $32$ nodes but higher energies for $64$ nodes, which is consistent with the simulator result.
As the problem size increases, the crossover between the bang-bang regime and the anneal regime generally occurs at longer operation times.
In other words, for a given operation time, the crossover from the anneal regime to the bang-bang regime appears as the system size increases.
This observation indicates that the pVSQA is more effective than pQA when the problem size is large.
For QKPs, the Powell optimizer fails to produce higher-quality solutions than pQA.
Fig.~\ref{fig:sdep} reveals that the ranges of $R_{50}^{\mathrm{ave}}$, representing the difference between the maximum and the minimum values, in the rescaled unit for QKPs are smaller than those for GPPs.
We hypothesize that the smaller ranges contribute to the relatively poor performances of the Powell optimizer for QKPs.
Even in this situation, the grid-search optimizer consistently obtains lower-energy solutions.
It is imperative to develop a classical optimizer that efficiently finds the optimal variational parameters on quantum annealers.

The gate-based quantum device demonstrates that pVSQA outperforms the conventional QAOA.
However, the experiments in this study are limited to a small-sized problem due to our restricted execution time on quantum devices.
In the future, the pVSQA performance should be evaluated on large-sized problems.

\section{Conclusion}~\label{Sec:Conclusion}
\hspace{-1mm}We proposed pVSQA to efficiently solve constrained COPs.
We provided a sufficient condition for constrained COPs to apply pVSQA based on a greedy post-processing algorithm. 
To demonstrate its feasibility for the state-of-the-art quantum annealer and gate-based quantum device, we applied it to two NP-hard COPs: GPP and QKP.
pVSQA on a simulator achieves a (near-)optimal performance within a predetermined operation time using a small number of variational parameters.
Then we experimentally verified the superior performances of pVSQA on real quantum devices compared to conventional quantum annealing with and without a post-processing and QAOA without a post-processing.

\section*{Acknowledgment}
This work was supported in part by JST CREST Grant Number JPMJCR19K4, Japan and KAKENHI Grant No.21K03391.
We acknowledge the use of IBM Quantum services for this work.
The views expressed are those of the authors, and do not reflect the official policy or position of IBM or the IBM Quantum team.

\bibliographystyle{IEEEtran}

\bibliography{DQA.bib}

\begin{thebibliography}{10}
\providecommand{\url}[1]{#1}
\csname url@samestyle\endcsname
\providecommand{\newblock}{\relax}
\providecommand{\bibinfo}[2]{#2}
\providecommand{\BIBentrySTDinterwordspacing}{\spaceskip=0pt\relax}
\providecommand{\BIBentryALTinterwordstretchfactor}{4}
\providecommand{\BIBentryALTinterwordspacing}{\spaceskip=\fontdimen2\font plus
\BIBentryALTinterwordstretchfactor\fontdimen3\font minus
  \fontdimen4\font\relax}
\providecommand{\BIBforeignlanguage}[2]{{%
\expandafter\ifx\csname l@#1\endcsname\relax
\typeout{** WARNING: IEEEtran.bst: No hyphenation pattern has been}%
\typeout{** loaded for the language `#1'. Using the pattern for}%
\typeout{** the default language instead.}%
\else
\language=\csname l@#1\endcsname
\fi
#2}}
\providecommand{\BIBdecl}{\relax}
\BIBdecl

\bibitem{kadowaki1998quantum}
T.~Kadowaki and H.~Nishimori, ``Quantum annealing in the transverse {I}sing
  model,'' \emph{Phys. Rev. E}, vol.~58, pp. 5355--5363, Nov 1998.

\bibitem{farhi2000quantum}
\BIBentryALTinterwordspacing
E.~Farhi, J.~Goldstone, S.~Gutmann, and M.~Sipser, ``Quantum computation by
  adiabatic evolution,'' \emph{arXiv:quant-ph/0001106}, 2000. [Online].
  Available: \url{https://arxiv.org/abs/quant-ph/0001106}
\BIBentrySTDinterwordspacing

\bibitem{morita2007convergence}
S.~Morita and H.~Nishimori, ``Convergence of quantum annealing with real-time
  schr{\"o}dinger dynamics,'' \emph{Journal of the Physical Society of Japan},
  vol.~76, no.~6, p. 064002, 2007.

\bibitem{johnson2011quantum}
M.~W. Johnson, M.~H.~S. Amin, S.~Gildert, T.~Lanting, F.~Hamze, N.~Dickson,
  R.~Harris, A.~J. Berkley, J.~Johansson, P.~Bunyk, E.~M. Chapple, C.~Enderud,
  J.~P. Hilton, K.~Karimi, E.~Ladizinsky, N.~Ladizinsky, T.~Oh, I.~Perminov,
  C.~Rich, M.~C. Thom, E.~Tolkacheva, C.~J.~S. Truncik, S.~Uchaikin, J.~Wang,
  B.~Wilson, and G.~Rose, ``Quantum annealing with manufactured spins,''
  \emph{Nature}, vol. 473, no. 7346, pp. 194--198, 2011.

\bibitem{maezawa2019toward}
M.~Maezawa, G.~Fujii, M.~Hidaka, K.~Imafuku, K.~Kikuchi, H.~Koike, K.~Makise,
  S.~Nagasawa, H.~Nakagawa, M.~Ukibe, and S.~Kawabata, ``Toward practical-scale
  quantum annealing machine for prime factoring,'' \emph{Journal of the
  Physical Society of Japan}, vol.~88, no.~6, p. 061012, 2019.

\bibitem{kordzanganeh2023benchmarking}
M.~Kordzanganeh, M.~Buchberger, B.~Kyriacou, M.~Povolotskii, W.~Fischer,
  A.~Kurkin, W.~Somogyi, A.~Sagingalieva, M.~Pflitsch, and A.~Melnikov,
  ``Benchmarking simulated and physical quantum processing units using quantum
  and hybrid algorithms,'' \emph{Advanced Quantum Technologies}, vol. n/a, no.
  n/a, p. 2300043, 2023.

\bibitem{king2022coherent}
A.~D. King, S.~Suzuki, J.~Raymond, A.~Zucca, T.~Lanting, F.~Altomare, A.~J.
  Berkley, S.~Ejtemaee, E.~Hoskinson, S.~Huang, E.~Ladizinsky, A.~J.~R.
  MacDonald, G.~Marsden, T.~Oh, G.~Poulin-Lamarre, M.~Reis, C.~Rich, Y.~Sato,
  J.~D. Whittaker, J.~Yao, R.~Harris, D.~A. Lidar, H.~Nishimori, and M.~H.
  Amin, ``Coherent quantum annealing in a programmable 2,000 qubit {I}sing
  chain,'' \emph{Nature Physics}, vol.~18, no.~11, pp. 1324--1328, 2022.

\bibitem{matsuura2021variationally}
S.~Matsuura, S.~Buck, V.~Senicourt, and A.~Zaribafiyan, ``Variationally
  scheduled quantum simulation,'' \emph{Phys. Rev. A}, vol. 103, p. 052435, May
  2021.

\bibitem{farhi2014quantum}
E.~Farhi, J.~Goldstone, and S.~Gutmann, ``A quantum approximate optimization
  algorithm,'' \emph{arXiv:1411.4028}, 2014.

\bibitem{zhou2020quantum}
\BIBentryALTinterwordspacing
L.~Zhou, S.-T. Wang, S.~Choi, H.~Pichler, and M.~D. Lukin, ``Quantum
  approximate optimization algorithm: Performance, mechanism, and
  implementation on near-term devices,'' \emph{Phys. Rev. X}, vol.~10, p.
  021067, Jun 2020. [Online]. Available:
  \url{https://link.aps.org/doi/10.1103/PhysRevX.10.021067}
\BIBentrySTDinterwordspacing

\bibitem{susa2021variational}
\BIBentryALTinterwordspacing
Y.~Susa and H.~Nishimori, ``Variational optimization of the quantum annealing
  schedule for the {{L}}echner-{{H}}auke-{{Z}}oller scheme,'' \emph{Phys. Rev.
  A}, vol. 103, p. 022619, Feb 2021. [Online]. Available:
  \url{https://link.aps.org/doi/10.1103/PhysRevA.103.022619}
\BIBentrySTDinterwordspacing

\bibitem{cote2023diabatic}
J.~C{\^o}t{\'e}, F.~Sauvage, M.~Larocca, M.~Jonsson, L.~Cincio, and T.~Albash,
  ``Diabatic quantum annealing for the frustrated ring model,'' \emph{Quantum
  Science and Technology}, vol.~8, no.~4, p. 045033, oct 2023.

\bibitem{finvzgar2023designing}
J.~R. Fin{\v{z}}gar, M.~J. Schuetz, J.~K. Brubaker, H.~Nishimori, and H.~G.
  Katzgraber, ``Designing quantum annealing schedules using {{B}}ayesian
  optimization,'' \emph{arXiv:2305.13365}, 2023.

\bibitem{lucas2014Ising}
A.~Lucas, ``Ising formulations of many {NP} problems,'' \emph{Frontiers in
  Physics}, vol.~2, p.~5, 2014.

\bibitem{shirai2022multi}
T.~Shirai and N.~Togawa, ``Multi-spin-flip engineering in an {{I}}sing
  machine,'' \emph{IEEE Transactions on Computers}, pp. 1--1, 2022.

\bibitem{kuramata2021larger}
M.~Kuramata, R.~Katsuki, and K.~Nakata, ``Larger sparse quadratic assignment
  problem optimization using quantum annealing and a bit-flip heuristic
  algorithm,'' in \emph{2021 IEEE 8th International Conference on Industrial
  Engineering and Applications (ICIEA)}, 2021, pp. 556--565.

\bibitem{fukada2021three}
K.~Fukada, M.~Parizy, Y.~Tomita, and N.~Togawa, ``A three-stage annealing
  method solving slot-placement problems using an {{I}}sing machine,''
  \emph{IEEE Access}, vol.~9, pp. 134\,413--134\,426, 2021.

\bibitem{glover2018tutorial}
F.~Glover, G.~Kochenberger, and Y.~Du, ``A tutorial on formulating and using
  {{QUBO}} models,'' \emph{arXiv:1811.11538}, 2018.

\bibitem{shirai2023spin}
T.~Shirai and N.~Togawa, ``Spin-variable reduction method for handling linear
  equality constraints in {I}sing machines,'' \emph{IEEE Transactions on
  Computers}, vol.~72, no.~8, pp. 2151--2164, 2023.

\bibitem{tanahashi2019application}
K.~Tanahashi, S.~Takayanagi, T.~Motohashi, and S.~Tanaka, ``Application of
  {I}sing machines and a software development for {I}sing machines,''
  \emph{Journal of the Physical Society of Japan}, vol.~88, no.~6, p. 061010,
  2019.

\bibitem{barkoutsos2020improving}
\BIBentryALTinterwordspacing
P.~K. Barkoutsos, G.~Nannicini, A.~Robert, I.~Tavernelli, and S.~Woerner,
  ``Improving {V}ariational {Q}uantum {O}ptimization using {CV}a{R},''
  \emph{{Quantum}}, vol.~4, p. 256, Apr. 2020. [Online]. Available:
  \url{https://doi.org/10.22331/q-2020-04-20-256}
\BIBentrySTDinterwordspacing

\bibitem{dezvalle2021quantum}
\BIBentryALTinterwordspacing
P.~D\'{\i}ez-Valle, D.~Porras, and J.~J. Garc\'{\i}a-Ripoll, ``Quantum
  variational optimization: The role of entanglement and problem hardness,''
  \emph{Phys. Rev. A}, vol. 104, p. 062426, Dec 2021. [Online]. Available:
  \url{https://link.aps.org/doi/10.1103/PhysRevA.104.062426}
\BIBentrySTDinterwordspacing

\bibitem{brady2021optimal}
\BIBentryALTinterwordspacing
L.~T. Brady, C.~L. Baldwin, A.~Bapat, Y.~Kharkov, and A.~V. Gorshkov, ``Optimal
  protocols in quantum annealing and quantum approximate optimization algorithm
  problems,'' \emph{Phys. Rev. Lett.}, vol. 126, p. 070505, Feb 2021. [Online].
  Available: \url{https://link.aps.org/doi/10.1103/PhysRevLett.126.070505}
\BIBentrySTDinterwordspacing

\bibitem{stefan2012heuristic}
S.~Edelkamp and S.~Schr{\"o}dl, ``Chapter 13 - constraint search,'' in
  \emph{Heuristic Search}, S.~Edelkamp and S.~Schr{\"o}dl, Eds.\hskip 1em plus
  0.5em minus 0.4em\relax San Francisco: Morgan Kaufmann, 2012, pp. 571--631.

\bibitem{pisinger2007quadratic}
D.~Pisinger, ``The quadratic knapsack problem---a survey,'' \emph{Discrete
  Applied Mathematics}, vol. 155, no.~5, pp. 623--648, 2007.

\bibitem{yoshimura2022qubo}
T.~Yoshimura, T.~Shirai, M.~Tawada, and N.~Togawa, ``{QUBO} matrix distorting
  method for consumer applications,'' in \emph{2022 IEEE International
  Conference on Consumer Electronics (ICCE)}, 2022, pp. 01--06.

\bibitem{mukasa2021scalable}
Y.~Mukasa, Y.~Matsuda, S.~Tanaka, and N.~Togawa, ``Scalable algorithms for
  capacitated vehicle routing problem using {I}sing machines,'' in
  \emph{Adiabatic Quantum Computing Conference 2021}, 2021.

\bibitem{tamura2021performance}
K.~Tamura, T.~Shirai, H.~Katsura, S.~Tanaka, and N.~Togawa, ``Performance
  comparison of typical binary-integer encodings in an {{I}}sing machine,''
  \emph{IEEE Access}, vol.~9, pp. 81\,032--81\,039, 2021.

\bibitem{patvardhan2015solving}
C.~Patvardhan, S.~Bansal, and A.~Srivastav, ``Solving the 0--1 quadratic
  knapsack problem with a competitive quantum inspired evolutionary
  algorithm,'' \emph{Journal of Computational and Applied Mathematics}, vol.
  285, pp. 86--99, 2015.

\bibitem{fixstars}
\BIBentryALTinterwordspacing
Fixstars {A}mplify {AE}. [Online]. Available:
  \url{https://amplify.fixstars.com/en/engine}
\BIBentrySTDinterwordspacing

\bibitem{scipy}
P.~Virtanen, R.~Gommers, T.~E. Oliphant, M.~Haberland, T.~Reddy, D.~Cournapeau,
  E.~Burovski, P.~Peterson, W.~Weckesser, J.~Bright, S.~J. {van der Walt},
  M.~Brett, J.~Wilson, K.~J. Millman, N.~Mayorov, A.~R.~J. Nelson, E.~Jones,
  R.~Kern, E.~Larson, C.~J. Carey, {\.I}.~Polat, Y.~Feng, E.~W. Moore,
  J.~{VanderPlas}, D.~Laxalde, J.~Perktold, R.~Cimrman, I.~Henriksen, E.~A.
  Quintero, C.~R. Harris, A.~M. Archibald, A.~H. Ribeiro, F.~Pedregosa, P.~{van
  Mulbregt}, and {SciPy 1.0 Contributors}, ``{{SciPy} 1.0: Fundamental
  Algorithms for Scientific Computing in Python},'' \emph{Nature Methods},
  vol.~17, pp. 261--272, 2020.

\bibitem{wiersema2020exploring}
\BIBentryALTinterwordspacing
R.~Wiersema, C.~Zhou, Y.~de~Sereville, J.~F. Carrasquilla, Y.~B. Kim, and
  H.~Yuen, ``Exploring entanglement and optimization within the hamiltonian
  variational ansatz,'' \emph{PRX Quantum}, vol.~1, p. 020319, Dec 2020.
  [Online]. Available:
  \url{https://link.aps.org/doi/10.1103/PRXQuantum.1.020319}
\BIBentrySTDinterwordspacing

\bibitem{boothby2020next}
K.~Boothby, P.~Bunyk, J.~Raymond, and A.~Roy, ``Next-generation topology of
  {{D}}-wave quantum processors,'' in \emph{D-Wave Technical Report Series},
  no. 14-1026A-C, 2019.

\bibitem{ibmq2021}
\BIBentryALTinterwordspacing
{IBM} {Q}uantum. [Online]. Available: \url{https://quantum-computing.ibm.com}
\BIBentrySTDinterwordspacing

\bibitem{choi2008minor}
V.~Choi, ``Minor-embedding in adiabatic quantum computation: I. the parameter
  setting problem,'' \emph{Quantum Information Processing}, vol.~7, no.~5, pp.
  193--209, 2008.

\bibitem{choi2011minor}
------, ``Minor-embedding in adiabatic quantum computation: {II}.
  minor-universal graph design,'' \emph{Quantum Information Processing},
  vol.~10, no.~3, pp. 343--353, 2011.

\bibitem{hadfield2019from}
\BIBentryALTinterwordspacing
S.~Hadfield, Z.~Wang, B.~O'Gorman, E.~G. Rieffel, D.~Venturelli, and R.~Biswas,
  ``From the quantum approximate optimization algorithm to a quantum
  alternating operator ansatz,'' \emph{Algorithms}, vol.~12, no.~2, 2019.
  [Online]. Available: \url{https://www.mdpi.com/1999-4893/12/2/34}
\BIBentrySTDinterwordspacing

\bibitem{qiskit2023}
{Qiskit contributors}, ``Qiskit: An open-source framework for quantum
  computing,'' 2023.

\end{thebibliography}

\EOD

\end{document}